# The Soreq Applied Research Accelerator Facility (SARAF) – Overview, Research Programs and Future Plans

*Israel* Mardor[1,2*], *Ofer* Aviv[2], *Marilena* Avrigeanu[3], *Dan* Berkovits[2], *Adi* Dahan[2], *Timo* Dickel[4,5], *Ilan* Eliyahu[2], *Moshe* Gai[6], *Inbal* Gavish-Segev[2], *Shlomi* Halfon[2], *Michael* Hass[7a] *Tsviki* Hirsh[2], *Boaz* Kaiser[2], *Daniel* Kijel[2], *Arik* Kreisel[2], *Yonatan* Mishnayot[2,8], *Ish* Mukul[7], *Ben* Ohayon[8], *Michael* Paul[8], *Amichay* Perry[2], *Hitesh* Rahangdale[8], *Jacob* Rodnizki[2], *Guy* Ron[8], *Revital* Sasson-Zukran[2], *Asher* Shor[2], *Ido* Silverman[2], *Moshe* Tessler[8], *Sergey* Vaintraub[2], and *Leo* Weissman[2].

[1]School of Physics and Astronomy, Tel Aviv University, Tel Aviv 69978, Israel
[2]Soreq Nuclear Research Center, Yavne 81800, Israel
[3]Horia Hulubei National Institute for Physics and Nuclear Engineering, P.O. Box MG-6, R-077125 Bucharest-Magurele, Romania
[4]GSI Helmholtzzentrum für Schwerionenforschung GmbH, 64291 Darmstadt, Germany
[5]II.Physikalisches Institut, Justus-Liebig-Universität Gießen, 35392 Gießen, Germany
[6]LNS at Avery Point, University of Connecticut, Groton, Connecticut 06340-6097, USA
[7]The Weizmann Institute of Science, Rehovot 76100, Israel
[8]Racah Institute of Physics, The Hebrew University, Jerusalem 91904, Israel

**Abstract.** The Soreq Applied Research Accelerator Facility (SARAF) is under construction in the Soreq Nuclear Research Center at Yavne, Israel. When completed at the beginning of the next decade, SARAF will be a user facility for basic and applied nuclear physics, based on a 40 MeV, 5 mA CW proton/deuteron superconducting linear accelerator. Phase I of SARAF (SARAF-I, 4 MeV, 2mA CW protons, 5 MeV 1mA CW deuterons) is already in operation, generating scientific results in several fields of interest. The main ongoing program at SARAF-I is the production of 30 keV neutrons and measurement of Maxwellian Averaged Cross Sections (MACS), important for the astrophysical s-process. The world leading Maxwellian epithermal neutron yield at SARAF-I ($5 \times 10^{10}$ epithermal neutrons/sec), generated by a novel Liquid-Lithium Target (LiLiT), enables improved precision of known MACSs, and new measurements of low-abundance and radioactive isotopes. Research plans for SARAF-II span several disciplines: Precision studies of beyond-Standard-Model effects by trapping light exotic radioisotopes, such as $^6$He, $^8$Li and $^{18,19,23}$Ne, in unprecedented amounts (including meaningful studies already at SARAF-I); extended nuclear astrophysics research with higher energy neutrons, including generation and studies of exotic neutron-rich isotopes relevant to the rapid (r-) process; nuclear structure of exotic isotopes; high energy neutron cross sections for basic nuclear physics and material science research, including neutron induced radiation damage; neutron based imaging and therapy; and novel radiopharmaceuticals development and production. In this paper we present a technical overview of SARAF-I and II, including a description of the accelerator and its irradiation targets; a survey of existing research programs at SARAF-I; and the research potential at the completed facility (SARAF-II).

---

[a] Deceased

*
Corresponding author: mardor@tauex.tau.ac.il

# 1 Introduction

Even after more than a century of research, major aspects of nuclear physics still remain unknown, especially away from the valley of stability, or those that require precise measurement of ultra-rare phenomena. Exploring this terra-incognita may shed new light on the genesis of elements in the universe, and may provide excellent probes to physics beyond the Standard Model of elementary particles.

New medium-high energy ion accelerators have been built or are under construction around the world in order to address these scientific challenges. Soreq Nuclear Research Center (SNRC) is constructing the Soreq Applied Research Accelerator Facility (SARAF), a user facility that will be based on a state-of-the-art light ion (protons and deuterons), medium energy (35 and 40 MeV) high CW current (5 mA) accelerator, to be completed by the beginning of the next decade. These cutting-edge specifications, with SARAF's unique liquid lithium target technology, can make SARAF competitive with the world's most powerful deuteron, proton and fast neutron sources.

Applied research and development at SARAF will seek to contribute to new nuclear medical treatments and pharmaceuticals, material characterization for more efficient and safer fission reactors and future fusion reactors, and accelerator-based neutron radiography, tomography and diffraction studies, currently available mainly in research reactors.

In Section 2 we describe the SARAF accelerator (the existing SARAF-I and the designed SARAF II), existing and planned high power irradiation targets, and the current and further required infrastructure. In Section 3 we report on the ongoing SARAF-I scientific research program, in Section 4 we elaborate on the potential and research plans of SARAF-II, and we conclude and outlook in Section 5.

# 2 Facility Description

In this Section we describe the accelerator and the irradiation targets of SARAF. We emphasize the reasoning behind the facility's chosen technologies and specifications.

## 2.1 The Accelerator

In Table 1 we present the accelerator top-level requirements and beam specifications, which will enable SARAF-II to be a world-leading high flux source of fast and high energy neutrons, and of radioisotopes for biomedical applications and basic science.

**Table 1**. SARAF-II beam top-level requirements

| Parameter | Value | Comment |
|---|---|---|
| Ion Species | Protons/Deuterons | M/q ≤ 2 |
| Energy Range | 5 – 40 MeV deuterons<br>5 – 35 MeV protons | Variable energy |
| Current Range | 0.04 – 5 mA | CW (and pulsed) |
| Operation | 6000 hours/year | |
| Maintenance | Hands-On | Low beam loss |

The demands for variable energy, proton and deuteron currents in the mA range and low beam loss that allows safe hands-on maintenance can be met only by a superconducting (SC) RF linear accelerator (linac). Independently phased RF cavities are needed for variable energy at the exit of the linac and for efficient acceleration of more than one ion species. Superconducting RF cavities are needed mainly for high beam power and large beam apertures that lessen beam loss.

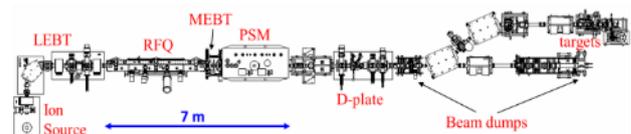

**Figure 1**. Layout of the accelerator at SARAF-I

An accelerator that covers all these requirements was not available when SARAF was founded, at the beginning of the century. Therefore, it was then decided to divide the accelerator construction into two phases. SARAF-I was built to test and characterize the required novel technologies and SARAF-II will be the complete accelerator with its final specifications (Table 1).

The SARAF-I linac (Figure 1) comprises a 20 keV/u, 7 mA, Electron Cyclotron Resonance (ECR) ion source, a Low Energy Beam Transport (LEBT), a 176 MHz, 1.5 MeV/u four-rod Radio-Frequency Quadrupole (RFQ), a short Medium Energy Beam Transport (MEBT) and a Prototype Superconducting Module (PSM) housing six Half Wave Resonators (HWR). SARAF-I is in operation since 2010, routinely accelerating up to 2 mA continuous-wave (CW) protons up to 4 MeV and 10% duty-factor deuterons up to 5.6 MeV.

A 13 m long High Energy Beam Transport (HEBT) line was built downstream of the SARAF-I accelerator, to characterize the linac's performance. Since 2011 the HEBT is being used to develop high intensity irradiation targets (Section 2.2) and run basic science experiments (Section 3).

The SARAF-II accelerator is currently under final design and first components have been tested at IRFU/CEA [1]. It consists of a new long MEBT, housing three normal conducting beam bunchers, and four SC modules each housing seven HWRs. The first two cryomodule HWRs are optimized for $\beta_0$=0.09 and the last two cryomodule HWRs are optimized for $\beta_0$=0.18.

The design, construction, commissioning, operation and use of SARAF-I components were described and published in [2, 3] and references therein. In the next three sections the main accelerator components are described, focusing on lessons learned at SARAF-I and realized in SARAF-II.

### 2.1.1 ECR Ion Source and LEBT

The SARAF ion source, built by RI GmbH [4], is a 2.45 GHz ECR, optimized for protons and deuterons. Ions are extracted by a 20 kV/u DC voltage to match the RFQ entrance velocity. The source plasma temperature and instabilities, the extraction voltage and the extraction electrode geometry define the initial beam quality. The beam is transported to the RFQ along the LEBT. A 90° bending magnet selects the ion mass to be transmitted in order to reduce potential beam loss downstream the RFQ. Three solenoid magnets focus the beam and provide appropriate matching to the RFQ.

Measured beam current after the bending magnet is 7 mA and higher. Typical rms normalized emittance measured in the slit and wire apparatus, about one meter upstream of the RFQ is $\varepsilon_{x,y;100\%}$=0.15 $\pi$·mm·mrad [5]. The beam is strongly focused towards the RFQ entrance by the third LEBT solenoid. It was shown that the beam reaches high space-charge neutralization along the LEBT, which reduces de-focusing. Nevertheless, the beam's rms emittance increases to an estimated value of about 0.20 $\pi$·mm·mrad at the RFQ entrance, due to an increased space charge force and leakage of the RFQ RF field [5].

A circular aperture downstream of the bending magnet may be used to cut transverse tails, reducing the beam intensity by a factor of up to 60. A beam chopper containing two deflecting-plates is used for further beam current attenuation during beam tuning and intensity ramping [3].

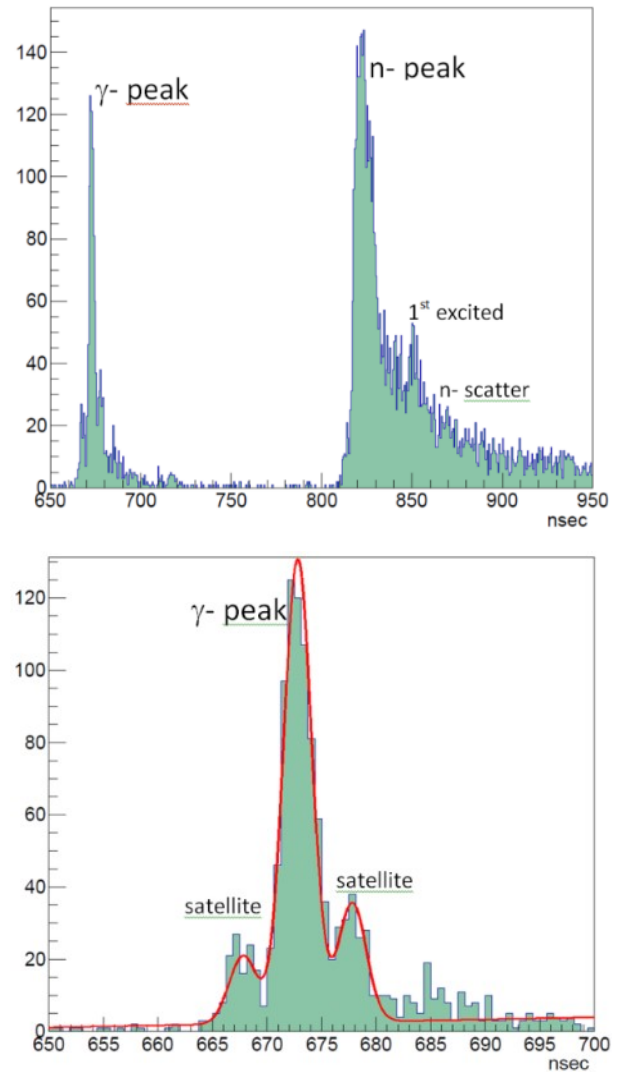

**Figure 2.** ToF measurements at SARAF with the fast chopper using 3.3 MeV protons on a thin LiF target. Spectrum is attained by a liquid scintillator 3.5 m away from the target. Top: overall ToF spectrum. The time differences between the neutron and gamma peaks, in conjunction with the target-detector distance, are consistent with neutron energies in the expected ~1 MeV range. Bottom: zoom in on gammas with a fit to the main and 2 satellite bunches. The peaks are ~ 5 ns apart, and the width of each peak is $\sigma$~1.3 ns, which is the ToF resolution.

Fast electronics enables operating this chopper as a single-bunch selector. It injects a 0.3 ns width (FWHM) beam pulse (shorter than the 5.6 ns RF period) into the RFQ, a key feature for neutron Time-of-Flight (ToF) measurements [6, 7]. This capability has been demonstrated at SARAF-I for protons at a pulse repetition rate of 40 kHz and is currently being upgraded for deuterons and a repetition rate of 200 kHz (duty factor $1.1\times10^{-3}$). In Figure 2 we show a ToF spectrum obtained by irradiating 3.3 MeV protons on a thin LiF target and detecting gammas and neutrons with a liquid scintillator.

One observes the gamma and neutron peaks, and also the separation between the main bunch and its preceding and following ones, parted by a period that is consistent with 5.6 ns. The ToF resolution is estimated by fitting each of the peaks, and is $\sigma$ ~ 1.3 ns.

### 2.1.2 The RFQ

The SARAF RFQ, built by RI GmbH and NTG GmbH [8] is of the four-rod (mini-vanes) type, containing 3.7 meters long electrodes. The RFQ was designed to accept the 20 keV/u DC beam emerging from the LEBT, focus it transversely and longitudinally at 176 MHz, and accelerate it as a CW beam up to 1.5 MeV/u at its exit.

The original RFQ design required an RF electrode voltage of 65 kV, corresponding to a power consumption of 250 kW, for deuteron acceleration (1.5 Kilpatrick's Criterion); however, repetitive sparking while approaching these values prevented CW deuteron acceleration [3].

Since 2010, the RFQ delivered CW beams of 2 mA protons at its nominal value, 1.5 MeV (32.5 kV electrode voltage, 62 kW power consumption), but was limited to a pulsed beam (10% duty cycle) for deuterons, at the nominal 3.0 MeV energy [3]. The RFQ could instantaneously hold msec pulses of the voltage required for deuterons acceleration, at up to 80% duty cycle. It was therefore assumed that the sparking was due to lack of heat removal at CW maximal power.

An RFQ upgrade program was initiated on 2014 in two parallel courses. First, the RF input line was split from one to two couplers. This removed the sparking problem in the original single coupler, which was identified at an early stage as a primary limit on RFQ total power [9].

Second, it was decided to reduce the power required for deuteron CW acceleration by reducing the deuterons (and protons) exit energy. Thus, a new rods modulation was designed for a reduced deuterons exit energy of 2.6 MeV (protons 1.3 MeV) which required a voltage of 56 kV and power consumption of 186 kW (keeping the same rods capacitance and maximal electric field of 1.5 Kilpatrick as in the original design). Detailed beam dynamics simulations, using the updated design of the accelerator, demonstrated that overall SARAF accelerator performance (Table 1) is not compromised by this reduction [10].

The new rods were installed in the SARAF RFQ in 2017, and after extensive conditioning and commissioning, CW deuteron beams of 1 mA were successfully accelerated through the RFQ to the required energy of 2.6 MeV, achieving a major milestone towards the realization of SARAF-II.

### 2.1.3 The Superconducting (SC) Linac

The SARAF-I SC linac, developed and built by RI GmbH, was designed as a demonstrator for the novel technologies necessary to construct SARAF-II. It is therefore composed of the following main components.

A 65 cm long MEBT with magnetic quadrupole lenses, steerers, vacuum pumps and limited beam diagnostics is used for transversal matching between the RFQ and the PSM.

A single, 4.5 K, 2.5 m long cryomodule (the PSM), houses six HWRs and three SC magnetic solenoids. The cavity optimal velocity ($\beta_0 = 0.09$) was determined as the minimal that could be fabricated, due to the narrow gaps in the acceleration region. This $\beta_0$ is still significantly higher than the RFQ exit ion velocity and thus imposes velocity mismatch and linac tuning difficulties. The first HWR in the PSM is used as a rebuncher to provide maximum longitudinal focusing.

The PSM at SARAF-I was the first one to demonstrate acceleration of:

- ions through HWR SC cavities
- ions through a separated vacuum SC cryomodule
- 2.1 mA CW variable energy protons beam

The two main drawbacks of the PSM design were the sensitivity of the HWRs to helium pressure variations and the limited relative alignment of the PSM components.

The HWRs were fabricated of 3 mm thick Nb sheets and installed in stainless steel liquid helium vessels. Due to the difference in heat expansion coefficients the Nb cavities were connected to the steel only around the beam ports. This design led to a measured RF frequency sensitivity of 60 Hz/mbar. Given the specified (and measured) helium pressure fluctuations of ±1.5 mbar, the HWR resonance frequency continuously varied up to ±90 Hz, outside of the cavity bandwidth of 130 Hz. These large offsets necessitated doubling of the RF amplifiers power to 4 kW (the maximum possible for the HWR couplers) in order to keep the HWR phases and voltages locked, but nonetheless limited the beam current and energy gain through the cavities. For more details see [2, 3] and references therein.

The SARAF-II HWRs are to be built with 2 mm thick Nb sheets and installed in Ti liquid helium vessels. Nb and Ti have very similar heat expansion coefficients so they will be connected at the beam ports and at four additional ports at the cavity's top and bottom ([1] and references therein). The calculated frequency sensitivity of this design is 7 and 3.3 Hz/mbar for the low- and high-beta HWRs, respectively (Table 2). With such a rigid cavity a fast tuner is no longer required and a slow tuner, firmer with respect to the SARAF-I HWR, will suffice.

As for relative component alignment in the PSM, the components small bore radii and lack of cold beam steering and diagnostic features inside the PSM forced a relative alignment criterion of better than ±0.2 mm. This was achieved at the factory PSM assembly, but was impossible to measure and guarantee after years of operation and several PSM cooling cycles. In the SARAF-II design, the HWR and solenoids beam bore radii were increased by 20% and 5%, respectively, and x-y steerers and cold beam-position-monitors (BPM) were added to each SC solenoid. Further, the ratio of solenoids to low $\beta_0$ HWRs in the linac lattice was increased from 1:2 to ~1:1 (Table 2), enabling better transverse beam control. This improved design enabled a more realistic and achievable relative alignment criterion of ±1 mm.

The operation experience of SARAF-I proved the applicability of HWR technology to light-ions high-intensity CW accelerators, and is therefore now used in the design and implementation of such accelerators world-wide, including the EURISOL driver [11], CADS injector-II [12], IFMIF/EVEDA [13], PXIE [14] and also for heavy ions drivers such as FRIB [15].

In Table 2, we summarize key SARAF linac parameters that changed from the original SARAF-I PSM design to the updated SARAF-II design.

**Table 2.** Key changed parameters between the original and updated SARAF linac designs. Note that some SARAF-II technical HWR specifications were not finalized in the original design, so they are marked 'NA'.

| | | Original | Updated |
|---|---|---|---|
| **RFQ** | | | |
| Rods Power | kW | 250 | 186 |
| Voltage | kV | 65 | 56 |
| Kilpatrick | | 1.54 | 1.52 |
| Exit energy | MeV/u | 1.50 | 1.27 |
| No. RF couplers | # | 1 | 2 |
| | | | |
| **HWR low beta ($\beta_0$=0.09)** | | | |
| $E_{peak}$ | MV/m | 25 | 35 |
| Bore dia. | mm | 30 | 36 |
| $E_{acc}(@E_{peak})$ | MV/m | 8.6 | 7.0 |
| $B_{peak}(@E_{peak})$ | mT | 53 | 66 |
| $\Delta V(@E_{peak})$ | MV | 0.845 | 0.95 |
| Sensitivity | Hz/mbar | 60 | 7 |
| | | | |
| **HWR high beta** | | 0.15 | 0.18 |
| $E_{peak}$ | MV/m | NA | 35 |
| Bore dia. | mm | 30 | 40 |
| $E_{acc}(@E_{peak})$ | MV/m | NA | 8.1 |
| $B_{peak}(@E_{peak})$ | mT | NA | 66 |
| $\Delta V(@E_{peak})$ | MV | 1.175 | 2.2 |
| Sensitivity | Hz/mbar | NA | 3.3 |
| | | | |
| **Solenoids** | | | |
| $Field_{peak}$ | T | 6.0 | 5.8 |
| Bore diameter | mm | 38 | 40 |
| x,y,z misalignment | mm | 0.2 | 1.0 |
| Steerer | | no | yes |
| Cold BPM | | no | yes |
| | | | |
| **40 MeV LINAC design** | | | |
| MEBT+SCL length | m | 21 | 25 |
| No. of low β HWR | # | 12 | 13 |
| No. of high β HWR | # | 32 | 14 |
| Total No. of HWR | # | 44 | 27 |
| No. of solenoids | # | 22 | 20 |

## 2.2 Irradiation Targets

The capabilities of new high intensity light-ion accelerators such as SARAF-II, SPIRAL-II [16], IFMIF\EVADA [13], and radio-pharmaceutical production cyclotrons such as BEST-70p [17] and IBA Cyclone-70 [18], pose new challenges to irradiation target development efforts. The combination of relatively low energy (40-70 MeV) and high current increases the volumetric power density deposited in the target and also the radiation damage to the target material.

The interaction of the particle beam with the target, both for radioisotope production and for neutron sources, generates very high density of thermal energy. The total beam power at SARAF-II will reach 200 kW. For some applications, this power is injected into small targets, generating power areal densities of ~1 to ~100 kW/cm². The design of such targets requires efficient heat removal techniques in order to preserve its integrity. In the following sub-sections, we present the research and development carried out in SARAF-I for this purpose.

### 2.2.1 Windowless Jet Liquid Lithium Target

The basic requirement for most SARAF-II applications is a high intensity compact neutron source that provides a high flux (luminosity) of neutrons at various energies. Hence, areal and volumetric power densities are high.

Several light target materials are available for producing variable energy neutrons by low energy light ions, including graphite, beryllium and lithium. Each material has advantages and requires specific technologies to form a robust target. Graphite has a high operating temperature and is implemented as a rotating target, using IR radiation for cooling [19, 20, 21]. Beryllium can be used as a static target with or without beam rastering [22, 23, 24], or in a rotating configuration [25]. Lithium can be solid for relatively low power [26, 27] or flowing liquid for high power, with or without a window [28, 29, 30, 31, 32].

Rotating targets require a large volume [19, 25], which leads to large neutron leakage if moderation is required. If necessary, both static and flowing liquid targets can be integrated into a moderator. Liquid targets are complex but are immune to radiation damage, blistering, and thermal stresses that limit the life time of static solid targets.

The advantages and drawbacks of these technologies are summarized in Table 3 and Table 4. Based on these insights, liquid lithium technology has been chosen to lead the development of high intensity neutron sources for SARAF-II applications.

**Table 3.** Target technologies and materials of several neutron and isotope sources in projects worldwide

| Design | Carbon | Beryllium | Lithium |
|---|---|---|---|
| **Static** | SPIRAL-I | LENS, SPES | Birmingham-BNCT, BINP |
| **Rotating** | SPIRAL-II, FRIB | ESS-Bilbao | Neutron Therapeutics Inc. |
| **Liquid** | X | X | IFMIF, SARAF |

**Table 4.** Advantages and drawbacks of the neutron and isotope source technologies presented in Table 3

| Design | Advantages | Drawbacks |
|---|---|---|
| **Static** | Simple, raster applicable | Localized heat flux (for small beam size), Radiation damage, Large neutron source |
| **Rotating** | Low thermal stress, Low radiation damage density | Complex mechanism, Reduced moderated neutron conversion efficiency |
| **Liquid** | Low thermal stress, Low radiation damage, Small neutron source | Complex system, complex operation and maintenance, only applicable light target is lithium |

The Liquid Lithium Target (LiLiT) at SARAF-I, schematically illustrated in Figure 3, consists of a film of liquid lithium (at ~200°C, above the lithium melting temperature of 180.5°C) force-flown at high velocity (up to 10 m/s) onto a concave thin stainless-steel wall. A rectangular-shaped nozzle just before the curved wall determines the film width and thickness (18 mm and 1.5 mm, respectively).

The liquid-lithium jet acts both as a neutron-producing target and as a beam dump, by removing, with fast transport, the thermal power generated by high intensity proton or deuteron beams. The target is bombarded directly on the windowless Li-vacuum interface. In recent years LiLiT has been used mainly to generate a quasi-Maxwellian neutron flux at $k_BT \sim 30$ keV. For this use, of specific concern is the volume power density, in the order of 1 MW/cm³, created by ~2 MeV protons in the Bragg peak area about 150 μm deep inside the lithium. The target was designed based on a thermal model, accompanied by a detailed calculation of the $^7Li(p,n)$ neutron yield, energy distribution, and angular distribution [33, 34, 35].

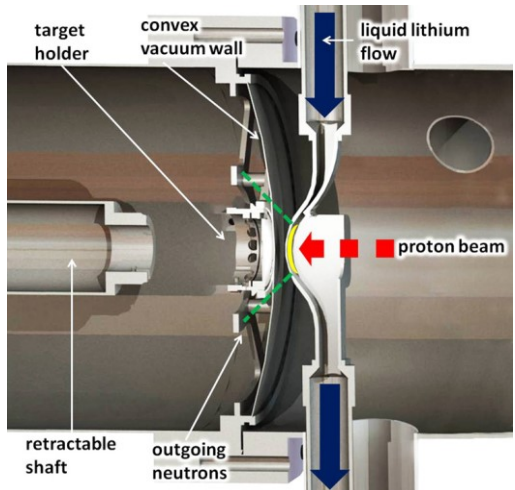

**Figure 3.** Cross section of LiLiT in the nozzle region. For details see [33]. The sample to be irradiated by the neutrons can be placed at about 6 mm from the target and is exposed there to a neutron flux of ~2×10¹⁰ n/s/mA. Reprinted with permission from [33].

LiLiT, installed in the SARAF-I accelerator beam line, is routinely irradiated with a high current proton beam. As of 2016, we have been operating it with a ~1.2 mA, ~1.926 MeV (2.3 kW) narrow proton beam ($\sigma_x =$ ~2.8 mm and $\sigma_y =$ ~3.8 mm) for periods of up to ten consecutive hours per irradiation. The proton energy was varied between 1.810 (below the neutron production threshold) and 1.940 MeV, determined by a time-of-flight measurement before and after each run. In such conditions the emitted neutron flux direction is forward, where a secondary experimental target is located and neutron reaction cross sections are studied.

In a typical experiment, after low-current beam tuning on a retractable Ta diagnostic plate located 5 cm upstream of the lithium nozzle, lithium flow is established and confirmed visually with a camera and its velocity is stabilized at ~2.4 m/s. The proton beam is set on the lithium film and the intensity is increased (within ~10 min) by increasing the accelerator beam duty factor until continuous-wave (CW) mode is reached. The temperatures of the lateral slabs on both sides of the nozzle are monitored by four thermocouple gauges and provide guidance for steering the beam to the center of the lithium flow in the horizontal direction. No disturbances or changes in the lithium film or lithium evaporation were observed during the experiments, in which vacuum was in the $10^{-5} – 10^{-6}$ mbar range.

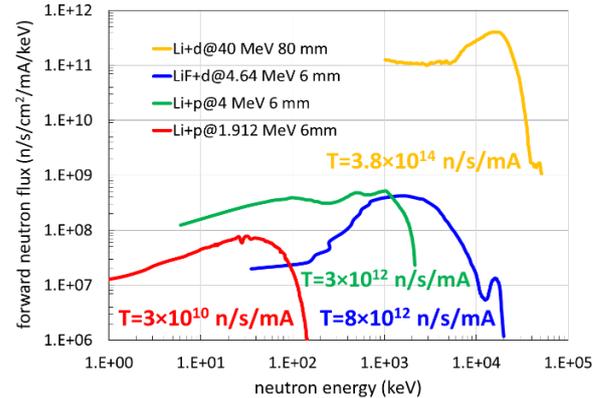

**Figure 4.** Neutron flux at θ=0° for four beam and thick Li target combinations. 80 and 6 mm are the distances from the neutron source. Red: simulated and measured at the SARAF-I s-process measurements with LiLiT [33]. Green: Simulated SimSARAF [34] output. Blue: simulated and measured at SARAF-I with a solid LiF target [37]. Orange: simulation based on measurements at Tohoku University [38]. T is the integral over 4π and neutron energy.

A secondary chamber downstream of the target acts both as a safety beam dump in case of accidental failure of the lithium flow, and as a convenient environment for positioning secondary targets. The curved wall of the nozzle allows activation foils to be positioned close (~6 mm) to the lithium surface, covering most of the outgoing neutron cone (target holder in Figure 3). Based on the activity of Au foils and the total charge (evaluated by the gamma dose and fission chamber measurements) the estimated total neutron rate was approximately 3×10¹⁰ n/s, forward collimated by the $^7Li(p,n)$ reaction kinematics. This flux is more than one order of magnitude larger than conventional neutron sources based on the near threshold $^7Li(p,n)$ reaction. The neutron energy spectrum was determined by simulations [33, 34]. Auto-radiographies of the abovementioned Au foils were also performed, revealing a narrow nearly Gaussian neutron distribution (FWHM ~ 15 mm) at the foil location [33].

Present and expected neutron intensities from proton and deuteron irradiation of liquid lithium targets at SARAF-I and II are presented in Figure 4 and Table 5.

**Table 5.** Present and expected proton and deuteron beams and neutron source intensities from Li targets at SARAF I and II. I+ means SARAF I + Phase I target room. All currents are CW.

| Phase (start year) | Beam | | | | Neutron Source | |
|---|---|---|---|---|---|---|
| | Proton | | Deuteron | | | |
| | E (MeV) | I (mA) | E (MeV) | I (mA) | E (MeV) | Rate (n/s) |
| **I (2013)** | 1.5-4 | 0.04-2 | 3-5.6 | 0.04-1.2 | 0.03-20 | 10¹¹ |
| **I+ (2018)** | 1.5-4 | 0.04-2 | 3-5.0 | 0.04-2 | 0.03-20 | 10¹³ |
| **II (2023)** | 5-35 | 0.04-5 | 5-40 | 0.04-5 | 0.03-55 | 10¹⁵ |

The neutron rates in Table 5 are for a thick target, producing a 'white spectrum' in the given energy range.

A quasi mono-energetic beam is attainable via thinner targets. In 2013-2017, the neutron flux was limited by the available concreted radiation shielding in the beam corridor, a limitation that will no longer exist once the SARAF-I target room is commissioned during 2018.

More information about LiLiT and its supporting systems (electro-magnetic pump, heat exchangers, control, etc.) can be found in [30, 31, 33, 35, 37].

The successful experience with LiLiT and its robust performance at SARAF-I made it the prime candidate to be the Thermal Neutron Source (TNS) for the Thermal Neutron Radiography (TNR) and diffraction systems at SARAF-II.

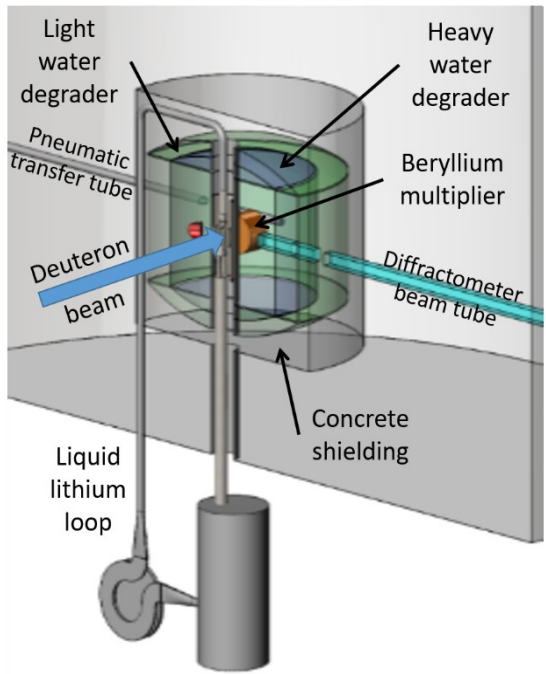

**Figure 5.** Schematic vertical cross-section of the liquid lithium target proposed for the SARAF II thermal neutron source. The liquid lithium is circulated in a vertical loop, and the deuteron beam (blue arrow) impinges upon it to generate neutrons, which are thermalized in the light and heavy water degraders and multiplied by the beryllium multiplier. For further details see Section 4.7.1 and Figure 30.

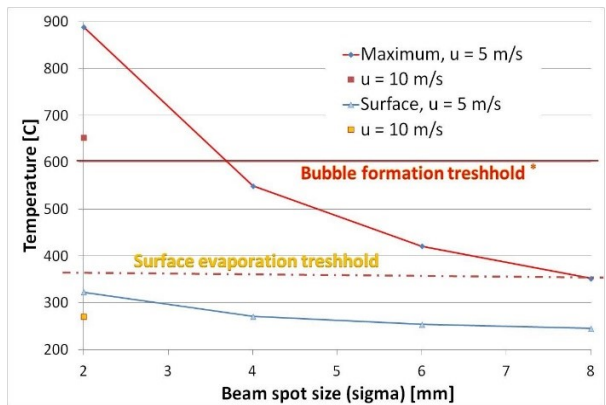

**Figure 6.** Lithium surface and maximum temperatures for 40 MeV/ 5mA deuterons versus beam size at two flow velocities (for 10 m/s only $\sigma$ = 2 mm is shown), compared to the estimated thresholds of bubble formation (*taken from [40]) and evaporation.

A preliminary design of a 200 kW liquid lithium target for a high intensity neutron source has been developed (Figure 5). Further information about the SARAF TNR system is given in Section 4.7.1.

The thermal design is based on a model presented in [35]. The model was built to estimate the flow velocity required to prevent excessive heating and evaporation of lithium in the beam spot area and possible upstream diffusion of vapors into the accelerator beam-line, and to eliminate boiling bubble formation inside the target, around the Bragg-peak area. Thus, the boiling temperature that depends on the fluid pressure, and the evaporation rate that increases exponentially with the surface temperature were studied as a function of beam spot size, together with the maximum temperature in the target, formed in the Bragg peak area (Figure 6).

The model indicates that for the SARAF TNS with a 200 kW, 40 MeV deuteron beam and a Gaussian particles distribution of $\sigma$ = 4 mm, the lithium jet velocity can be as low as 5 m/s. This will require a jet thickness of 25 mm and width of 50 mm. The estimated jet temperature distribution for these conditions is shown in Figure 7.

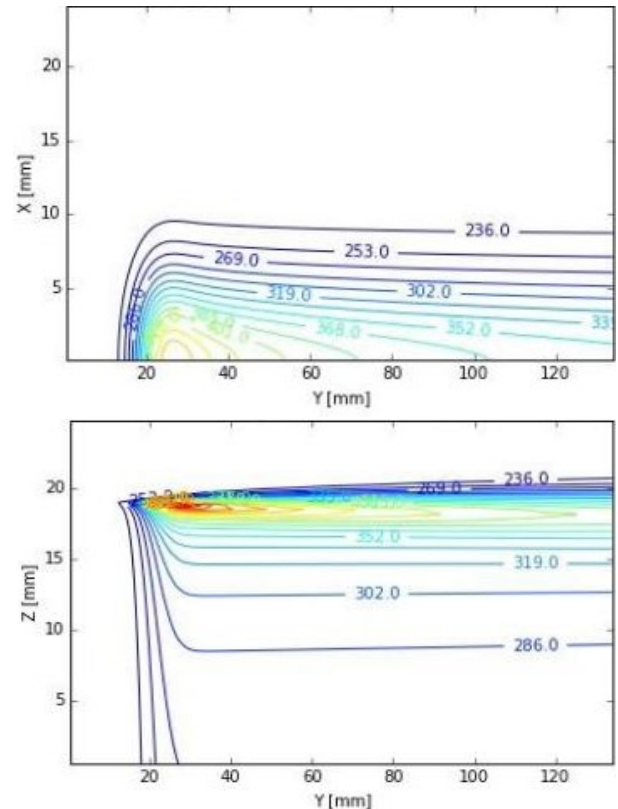

**Figure 7.** Simulated temperature distributions (°C) in a 25 mm thick lithium jet, flowing at 5 m/sec, irradiated by 40 MeV, 5 mA deuterons. Beam size is $\sigma$ = 4 mm. Y is the flow direction, X is perpendicular to it, Z is the beam direction. Beam hits the lithium jet at its transversal center (x = 0) and at y = 20 mm. Top: Distribution at Z = 19.1 mm. Bottom: Distribution at transversal center.

### 2.2.2 Encapsulated liquid and foil composite targets

For radio-pharmaceutical isotopes research, development and production, one usually uses encapsulated liquid and thin foil composite targets. Such targets are used in cyclotrons, where cooling is commonly performed by a high velocity jet of water that exerts approximately 20 bars on the target's back. This requires a high pressure pump and causes high mechanical load on the back of the irradiated capsule, whereas the front side of the target is exposed to the beam and the vacuum of the cyclotron.

It is essential that the irradiation target be kept small (implying a high heat flux), to generate a high specific activity product and for lowering of carrier material contamination of the radio-pharmaceutical target, and because for many cases the target material is an enriched, expensive isotope.

Several techniques can be used to overcome the high heat fluxes expected for these targets. An example is jet-impingement that has been shown to cope with heat fluxes up to 40 kW/cm², over an area of few square mm, using water as coolant [41]. Blackburn et al. used this technique to develop a high-power target for production of neutrons in accelerators [44]. Lienhard [43] provided a detailed review of the technique. Data of several candidate coolants for high heat flux cooling systems, including water, is given in Table 6. For example, Blackburn used gallium as a coolant of a BNCT neutron source [44] and Carpenter still advocates it [44].

**Table 6.** Physical parameters of several coolants

| Parameter | H₂O | Ga | GaIn | NaK | Hg |
|---|---|---|---|---|---|
| Composition | | | 77% Ga 22% In | 78% Na 22% K | |
| Melting T [°C] | 0 | 29.8 | 15.7 | -11.1 | -38.8 |
| Boiling T [°C] | 100 | 2205 | 2000 | 783.8 | 356.8 |
| Density [kg/m³] | 1000 | 6100 | 6280 | 872 | 13599 |
| Heat Capacity [J/kg·K] | 4181 | 373 | 326 | 1154 | 140 |
| Thermal Conductivity [W/m·K] | 0.61 | 28 | 41.8 | 25.3 | 7.8 |
| Viscosity [10⁻³ kg/m·s] | 0.855 | 1.96 | 1.69 | 0.468 | 1.55 |
| Kinematic Viscosity [10⁻⁸ m²/s] | 85.5 | 32 | 27 | 53.7 | 11.4 |
| Prandtl Number | 5.86 | 0.0261 | 0.0204 | 0.0213 | 0.0278 |

In the expected power and power density of SARAF-II, the pressure difference between the two sides of the target, together with the high target temperature, may rupture such a thin foil structure; therefore, the SARAF Target Development Group developed targets for high power accelerators [46]. In particular, the Group demonstrated the capability to remove very high heat flux from targets with limited mechanical impact on them, and concluded that reliable and affordable high heat flux cooling systems that are based on jet impingement can be designed and built [47]. These cooling systems can remove at least 4 kW/cm² with high velocity, high pressure water jets.

Cooling with liquid metal (e.g. gallium) jet-impingement may provide more effective cooling, up to an average heat flux of 8.4 kW/cm² with negligible pressure, as demonstrated in [47]. This challenging design was implemented successfully due to a custom electromagnetic pump that circulated the liquid gallium. A schematic drawing of one of the SARAF encapsulated liquid targets is presented in Figure 8.

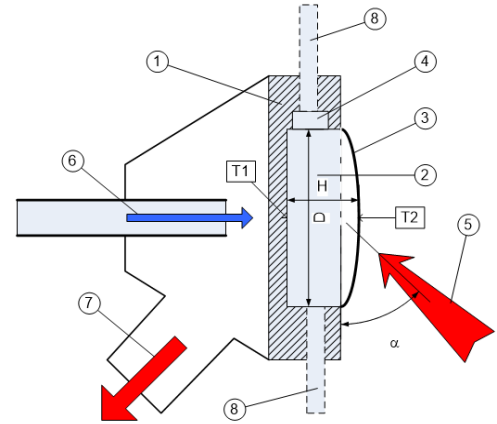

**Figure 8.** Conceptual drawing of the encapsulated liquid target. 1) capsule body, 2) irradiated material (natGa), 3) thin Havar window, 4) expansion volume, 5) particle beam, 6) cooling jet, 7) coolant outflow, 8) optional irradiated material ports.

The gallium target is usually encapsulated in niobium, which is used since gallium is very corrosive, especially at high temperatures, and attacks other metals and alloys.

The target is made of a molybdenum alloy (TZM) capsule enclosing the irradiated material with a thin HAVAR metallic window, which separates it from the beam line, and a strong heat conductive base that is cooled by a liquid metal impingement jet. TZM has high thermal conductivity and high strength up to high temperature. HAVAR foils are commonly used for FDG production targets due to their extremely high tension strength up to 500° C and above. Both are plated with a thin niobium layer to reduce corrosion [48].

The capsule may be sealed and used once, or equipped with ports for injecting and extracting the irradiated material. A small gas-filled expansion volume is provided to reduce pressure fluctuations within the liquid irradiated material during target temperature transients. The target is positioned at 45° to the beam axis in order to reduce the heat flux and increase the cooling area.

The limiting factors on the capsule mechanical design are stresses and radiation damage in the thin window. We designed and checked HAVAR and SS-316L foils for such a window. This window should withstand the internal pressure of the liquid gallium and is the warmest part of the capsule. It also suffers high radiation damage from the medium energy high flux particle beam.

More details on our work related to this type of irradiation targets can be found in [46, 47, 49, 50, 51, 52, 53].

### 2.2.3 High Power Beam Dumps

A high power accelerator requires an appropriate beam dump to contain its beam in a robust manner. The technology investigated during SARAF-I was of high power solid irradiation targets. Cooling of such targets is usually achieved by either jet-impingement, described in Section 2.2.2, or mini/micro-channels.

Micro-channels cooling [27] is advantageous as it can be integrated into the target and provide internal cooling.

The SARAF Target Development Group studied this technique as a possible design for the thermal neutron source of SARAF-II, and concluded that it can achieve high heat flux cooling in the range of 1 kW/cm² and above [54]. However, it was eventually decided that the liquid lithium target technique will be used for the thermal neutron facility (Section 4.7.1), and the micro-channel technique was since mainly used at SARAF-I for high power beam dump development.

Several generations of beam dumps, designed to handle a high power beam, were built and used at SARAF-I. The material of choice was mostly W (or W alloys) since no nuclear reactions with residual activity are expected with tungsten at proton and deuteron energies below 8 MeV, which was confirmed experimentally.

The first version was a bulk heavy-metal (HM 97% tungsten copper free) ConFlat (CF) 8" OD flange. This flange was machined out of HM produced by metal powder sintering. The HM flange was mounted between two stainless steel (SS) flanges, one on the beam vacuum front side and the other on the back side. Cooling water was driven at a pressure of 4 bar between the HM and the back SS flanges, using millimeter-size channels machined in the SS flange.

This beam dump was tested with a 6 kW 1.5 MeV proton beam. The heat removal was good, but at high beam power and high water pressure the vacuum pressure in the beam pipe increased rapidly due to hydrogen presence. It was assumed that the hydrogen originated from the cooling water through the HM, which became porous and transparent to hydrogen at these conditions. In addition, due to the difference in heat expansion coefficients between HM and SS, the three CF flanges were fractured after a few beam heating cycles.

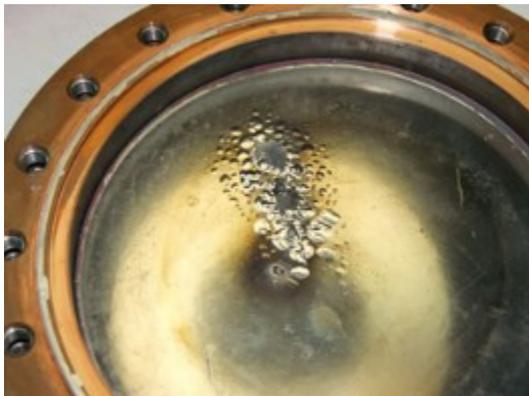

**Figure 9.** A 20 cm diameter tungsten beam dump, damaged after several hours of irradiation by 4 MeV / 1 mA protons. Extensive bubble formation and exfoliation is clearly visible. Reprinted with permission from [55].

The following beam dump version had a thin (250 µm) layer of tungsten or tungsten alloys brazed to a massive water cooled copper flange, It turned out that high heat fluxes were not the limiting factors in these cases, where the target size is quite large, but rather radiation damage and blistering due to gases accumulation in the target material (from beam particles or isotope production). In Figure 9 we show an example of a destroyed beam dump due to hydrogen blistering, merely after several hours of irradiation. In addition, the small amount of alloying elements was activated.

At high beam power, the bursting of such blisters erodes the beam dump surface (Figure 9), and rapidly increases the beam pipe vacuum pressure [56], which might cause accelerator interlock issues. In conclusion, the SARAF operational experience is that this beam dump concept is not satisfactory.

The diffusion rate of hydrogen in metals increases with temperature. Therefore, if the beam dump is operated at higher temperature, the threshold for blister formation increases (see Section 3.6). This understanding has been incorporated into the design of the third generation beam dump for SARAF-I, which comprised up to 81 pure (99.99%) tungsten pins inserted in holes within a water cooled copper base-plate [57, 58]. Each pin is 4 mm in diameter and 6 cm long. The pins are inserted to a depth of 2 cm.

The base-plate is embedded into a flange placed at the end of the beam pipe. It is tilted at a shallow angle with respect to the vertical direction and correspondingly the pins are tilted with respect to the direction of the incoming beam. This is in order to shadow the copper surface between the pins from the incoming beam and, hence, to minimize nuclear reactions during operation. Photographs of the tungsten-pins beam dump are shown in Figure 10.

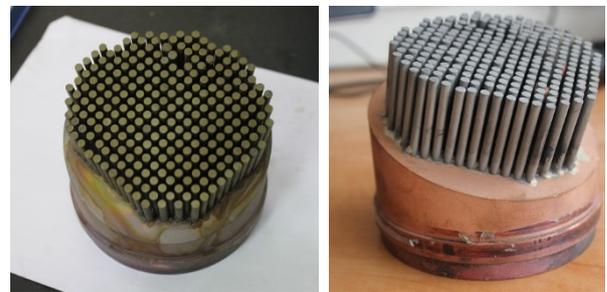

**Figure 10.** Photographs of the tungsten-pins beam dump, depicting the layout of the pins and their inclination with respect to the accelerator beam direction.

Primary removal of beam induced heat is by radiation from the pins, which reach high temperatures during irradiation; moreover, this high temperature ensures fast, efficient diffusion of the implanted hydrogen from the pins [59]. In addition, due to their production method (cold rolling), the tungsten pins have porous micro-channels along their axes, which further facilitate fast diffusion of implanted hydrogen.

This beam dump is operated rather successfully at SARAF-I using medium power beams for several years. However, due to the high temperature it is somewhat sensitive to over-focusing of the beam, which might melt the tungsten pins. Two methods were tested to spread the beam spot size across the pins surface; defocusing by a magnetic lens, and steering by a magnetic raster. Using the latter method, the tungsten-pin beam dump was operated successfully with a ~4 kW beam (~4 MeV, 1.5 mA, ~70% duty cycle), which was spread across the beam dump resulting in a beam heat flux of ~300 W/cm². Prompt and residual radiation were in the ~20 and ~4 mRem/hr range, respectively, and no blistering damage was observed. Further details about this beam dump and its testing procedures and results are given in [58].

Due to the overall sensitivity and power (and also power density) limit of solid beam dumps, for the 40 MeV, 200 kW SARAF-II we are considering a liquid metal beam dump, possibly made of gallium-indium, enabling a liquid target thickness of a few mm.

# 3 SARAF-I Research Programs

## 3.1 Nuclear Astrophysics – The s-Process

The $^7Li(p,n)^7Be$ reaction just above the $^7Li(p,n)$ threshold (Ep = 1.880 MeV) has been traditionally used to produce neutrons in the epithermal energy regime; the thick-target angle-integrated neutron yield is known to have an energy distribution similar to that of a flux of Maxwellian neutrons with $k_BT\sim25$ keV [60].

The LiLiT unique high epithermal neutron yield, $\sim$3– $5 \times 10^{10}$ n/s, is 30–50 times larger than in existing facilities based on the near-threshold $^7Li(p, n)$ reaction for neutron production. The neutron energy spectrum fits well a Maxwellian also with the broad energy spread of the SARAF-I proton beam ($\sigma \sim 20$ keV) [61]. This high yield was used in SARAF-I in recent years for Maxwellian-Averaged Cross Section (MACS) measurements of $^{94,96}Zr$ [33], $^{36,38}Ar$, $^{78, 80,84,86}Kr$, $^{136,138,140,142}Ce$, $^{147}Pm$, $^{171}Tm$ [62], $^{63}Cu$, $^{208}Pb$ [63] and $^{209}Bi$ [64]. The full program of MACS measurements at LiLiT is given in [62]. The high neutron intensity and flux of LiLiT will enable extending experimental investigations to small cross section, low-abundance and radioactive targets.

The operation of LiLiT is described briefly in Section 2.2.1. Further details about LiLiT operation and monitoring, and the design, performance and analysis of MACS experiments are given in [30, 33, 34, 35].

A review of LiLiT MACS studies, presenting experimental results and their significance to s-process nucleosynthesis research, is under preparation [65].

## 3.2 Nuclear Astrophysics – The primordial $^7Li$ problem

The disagreement of the predicted abundance of primordial $^7Li$ with the observed abundance is a longstanding problem in Big Bang Nucleosynthesis (BBN) theory. While BBN theory correctly predicts the relative abundances of $^2H/^1H$, $^3He/^1H$ and $^4He/^1H$ (that vary over four orders of magnitudes), it predicts the relative abundance of primordial $^7Li/^1H$ by a factor of approximately 3-4 larger than observed (approximately 4-5$\sigma$ discrepancy) [66].

It was proposed to measure at SARAF-I the direct destruction of $^7Be$ during (the first 10-15 minutes of) BBN via the $^7Be(n,\alpha)$ reaction with neutron of a Maxwellian energy distribution around 49.2 keV (consistent with the expected BBN energy 20-100 keV), in order to check whether this direct destruction of $^7Be$ competes with the $^7Be(n,p)^7Li(p,\alpha)$ reaction chain and thus explains the abovementioned discrepancy [67].

A $^7Be$ target of $\sim$4 GBq was produced in PSI [68] and delivered to SARAF-I. The target was positioned in the LiLiT experimental chamber and irradiated by high flux

49.2 keV quasi Maxwelian neutrons. The generated $\alpha$ particles (and also protons from $^7Be(n,p)$) were measured with CR39 nuclear track detectors. The analysis of this experiment is ongoing [69].

As can be seen in [70], there is direct $^7Be(n,\alpha)$ data only up to $E_n \sim 12.7$ keV, well below the BBN window stated above, and the cross section at higher energies is deduced from inverse measurements, down to $E_n = 228.6$ keV, well above the BBN window. Therefore, a follow-up measurement at the BBN window is proposed at SARAF-I, where the reaction will be measured with an array of diamond detectors [71]. Such detectors will enable measuring the two emerging alpha-particles from $^7Be(n,\alpha)$ in coincidence, and can withstand the SARAF-I high neutron flux.

## 3.3 Precision Measurements of Standard Model Parameters

Investigation of the kinematics of charged particles emerging from nuclear beta-decay is used to search for beyond-standard-model (BSM) physics in the weak sector (see Section 4.3 for more details). A precise measurement involves collection of many (few $10^7$) decays in a trap, which requires the production of large amounts of trappable, light, radioactive isotopes. Deep understanding of their decay scheme and various corrections are crucial for assessing systematic errors. Thus, our previous efforts (at The Weizmann Institute of Science (WIS) and at SARAF-I) and current (at SARAF-I) focus on improvement of the production and transport scheme, and measurement of nuclear physics observables pertaining to the decay.

Meaningful experiments, mainly measuring the β–ν correlation in the β-decay of $^6He$ and $^{23}Ne$, can be conducted with the energies and currents available at SARAF-I+ (Table 5).

### 3.3.1 Production of $^6He$

In the proposed scheme, the radioisotopes $^6He$ are produced via $^9Be(n,\alpha)^6He$, in which neutrons hit a hot (1,500°C) and porous BeO fiber disks target placed inside a UHV furnace. Due to the high temperature and porous (50%) nature of the target, the produced $^6He$ atoms effuse out efficiently and move to the measurement area. This concept was demonstrated at ISOLDE with spallation neutrons from impingement of 1.4 GeV protons on W, and up to $4.1\times10^{10}$ $^6He/\mu C$ of protons were extracted, an unprecedented and yet unmatched yield [72]. It was recently commissioned and successfully tested at WIS, using a 14 MeV neutron beam of flux $\sim2\times10^{10}$ n/sec from commercial D-T neutron generator [73].

In the WIS setup, the yielded $^6He$ atoms were transported to a decay measurement chamber. The transport line was about nine meters long and included a cryogenic pump for purification. The current system has a production probability of $\sim1.45\times10^{-4}$ $^6He$ per neutron and is $4.0\pm0.6\%$ efficient for the transport of $^6He$ from target to a measurement chamber. The background-subtracted electron spectrum obtained from the decay of $^6He$ is shown in Figure 11, along with a comparison to GEANT4 [74] simulations.

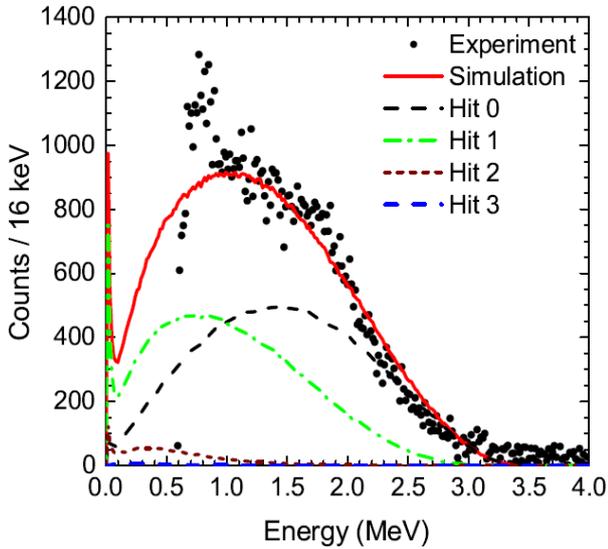

**Figure 11.** Experimental and GEANT4 simulated spectra of electrons from $^6$He beta-decay inside the EIBT. Black dots: Background-subtracted experimental spectrum. Solid red line: Normalized total simulated response. Other lines: simulated spectra of detected electrons after 0, 1, 2 and 3 hits of the vacuum chamber walls, as listed in the legend. Reprinted with permission from [73].

The observed spectrum is consistent with a simulation that contains a sum of electrons arriving directly from the $^6$He decay event, and those that hit the vacuum chamber walls up to three times before being detected. As such, it is an important validation of the large scale-up of the present target mass of BeO compared to the BeO target in the ISOLDE experiment [72] and bodes well for possibly even larger scale-ups that are planned for the implementation at SARAF-I.

The $^6$He production and measurement system, dubbed Weizmann Institute and SARAF Electrostatic Trap for Radio-Active Particles (WISETRAP), was recently transported to a new laboratory at SARAF, located above the SARAF target room (Figure 12).

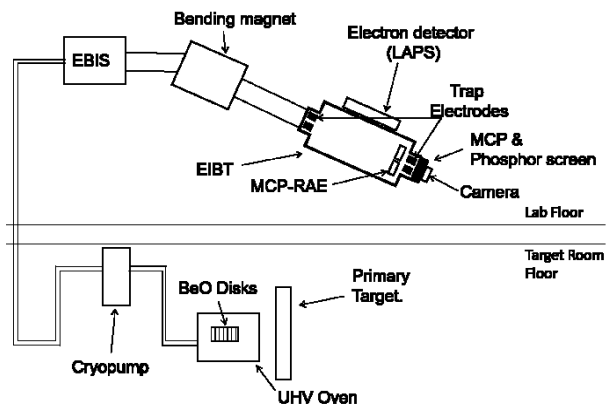

**Figure 12.** Schematic layout of WISETRAP at SARAF-I for $^6$He production and measurements. Primary target is LiF for SARAF-I and LiLiT for SARAF-II. For the production experiment at WIS, neutrons from a neutron generator hit the BeO disks, and produced He atoms that effused to a measurement chamber placed downstream of the cryopump.

Subsequent $^6$He production experiments are planned, using a solid LiF target for neutron production, to measure the production and transport efficiency of the setup to the Electrostatic Ion Beam Trap (EIBT), which will be eventually used to measure the $^6$He decay properties (Section 4.3.2).

With respect to the WIS setup, at SARAF-I we expect to increase the $^6$He production rate by at least two orders of magnitude due to an increase in the neutron flux, the size of targets and inclusion of more BeO disks, and to an increase the transport efficiency to nearly 90% by using larger diameter transmission pipes.

### 3.3.2 Production of $^{23}$Ne

The first unstable Ne isotope investigated at SARAF is $^{23}$Ne, as it is the most reachable, requiring neutrons of only a few MeV impinging upon a Na target (Table 9).

A $^{23}$Ne production experiment at SARAF I took place on 2017. It examined our ability to produce and transport large quantities of radioisotopes from the target to a science area removed from the high neutron radiation.

A lithium fluoride (LiF) target was mounted downstream of the accelerator. A deuteron beam of up to 10μA was accelerated to 5 MeV, hitting the LiF target and producing neutrons via Li(d,n). The neutrons then hit a 1.4 kg ground NaCl target heated to 573K, producing $^{23}$Ne atoms via $^{23}$Na(n,p)$^{23}$Ne. The $^{23}$Ne atoms diffused out the NaCl crystals, effused out the target and diffused to a measurement cell through a ten meters long, three inches thick hose. The cell was located in the accelerator service corridor.

Decay products of the $^{23}$Ne atoms that decayed in the measurement cell were measured by an HPGe detector, and thin and thick plastic scintillators, which were used as ΔE/E telescopes to identify electrons. The gamma spectrum measured via the HPGe detector is given in Figure 13 and clearly shows the 440, 1636, and 2076 keV gamma lines for decays to various exited states of $^{23}$Na (Figure 14). Quantitative analysis of this experiment is ongoing.

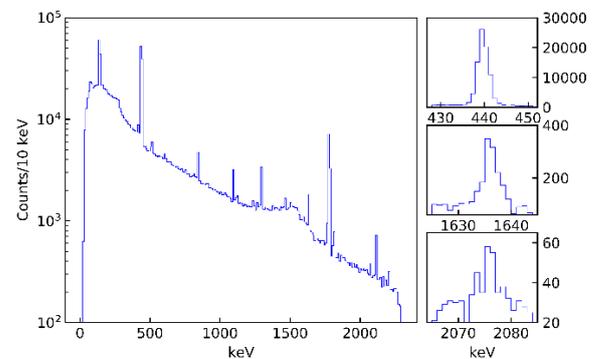

**Figure 13.** The measured gamma spectrum of $^{23}$Ne. The 440, 1636 and 2076 keV peaks are shown at insets (whose vertical axes are counts/kev) and pertain to excited states of $^{23}$Na (Figure 14). Activation lines of $^{56}$Mn (847, 1811 and 2113 keV) from the HPGe detector's stainless steel cell, and of $^{28}$Al (1779 keV) from the HPGe detector cap, are also observed. The measurement lasted 48 min.

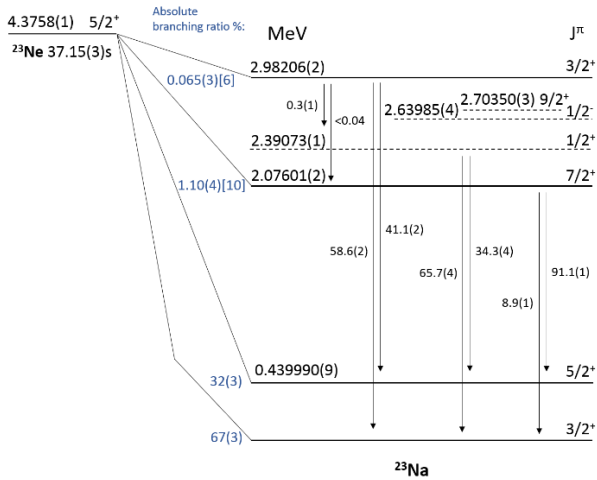

**Figure 14.** $^{23}$Na lower levels and $^{23}$Ne decay scheme. Based on data from [75, 77, 78, 79, 80, 81], not assuming Standard Model (SM) angular correlations. Allowed transitions are indicated along with the branching ratios. Branching to 2.076 and 2.982 MeV levels is estimated from gamma ray intensities and so is systematically dependent on the branching uncertainty to the first excited state.

A follow-up experiment is currently underway, which is aimed to measure various parameters that affect $^{23}$Ne production, such as target temperature and the branching ratio (BR) of $^{23}$Ne decay to the ground and first excited states (Figure 14). This experiment will be performed with a new disc shaped measurement cell. The measurement cell design is based on GEANT4 simulations, aimed at improving the BR measurement.

This BR was measured by Penning and Schmidt with uncertainty of 10% [75], and this large uncertainty was the leading contribution to the uncertainty in the most recent measurement of the β−ν correlation coefficient $a_{\beta\nu}$ made by Carlson in 1963 [76]. Thus, a precise determination of the above branching ratio is crucial for a more precise determination of $a_{\beta\nu}$.

These are the first steps in a series of experiments aimed at investigating the β-decay of $^{23}$Ne at SARAF, where the near future experiments at SARAF-I will yield ~$10^9$ atoms/sec, and consequent ones in SARAF-II will yield an unmatched quantity of as many as ~$10^{11}$ $^{23}$Ne atoms/sec (Table 9).

## 3.4 High Energy Neutron Production

Generating high energy neutrons at SARAF-I, with energies exceeding the beam energy, can be achieved by using exothermic reactions such as $^7$Li(d,n)$^8$Be, with a Q-value of 15.03 MeV. To utilize this reaction with beam power up to a few kW, a solid Lithium Fluoride Thick Target (LiFTiT) [82] was developed.

LiFTiT consists of a 250 μm thick LiF crystal, 30 mm in diameter, attached to a copper backing that comprises an inner water cooling loop. To guarantee efficient cooling of the copper backing during dissipation of the few kW deuteron beam power, water is pressurized through 300 μm micro-channels [27, 54].

The copper backing can be mounted on flanges of different diameters, which have to be electrically insulated to enable beam current measurement during experiments. Due to its compact design, LiFTiT facilitates optimal collection of neutrons inside a secondary target, in view of our two-stage scheme for light exotic isotope production (Section 4.3.1). A CAD drawing of LiFTiT is shown in Figure 15.

The target was operated at SARAF-I with a pulsed deuteron beam of $E_d$ = 4.64 MeV, at a frequency of 10 Hz, a pulse time window width of 1 msec and about 100 μC per pulse. The average deuteron current on-target during the experiment was measured to be around 1 μA.

The LiFTiT neutron spectrum was measured by two independent methods; activation foil irradiation, a technique that we used in the $^6$He production experiment at ISOLDE (Section 3.3.1 and [72], $E_n$ up to 100 MeV), but with foils that are appropriate to lower neutron energy range (up to 20 MeV), and in-beam measurement by an NE-213 liquid scintillator (LS) [83] that is known for its good efficiency for fast-neutrons and good separation between neutrons and gammas. Nonetheless, during the experiment we placed a 5 cm thick lead shield in front of the LS, in order to block most of the intense prompt gamma emission.

Neither of these methods provides a direct measurement of a neutron spectrum; as a result, the neutron spectrum has to be "unfolded" from the raw measurement, by various techniques that are described in detail in [37]. In general, the unfolding process requires prior theoretical estimation of the neutron spectrum, which is provided by Monte-Carlo simulations of the specific experimental setup.

In the analysis of the above experiment, we initialized the unfolding of the LS spectra via a 'guess' spectrum generated by MCUNED [84], which is a special code for simulating neutron spectra following deuteron induced reactions, in conjunction with the LS response functions calculated by the code NRESP7 [85]. Due to possible neutron scattering in the abovementioned 5 cm lead shield and in the surrounding thick concrete walls, and the fact that the LS response functions were not calibrated experimentally, we considered the resultant unfolded LS spectrum merely as a 'modified guess' spectrum for the activation foils measurement.

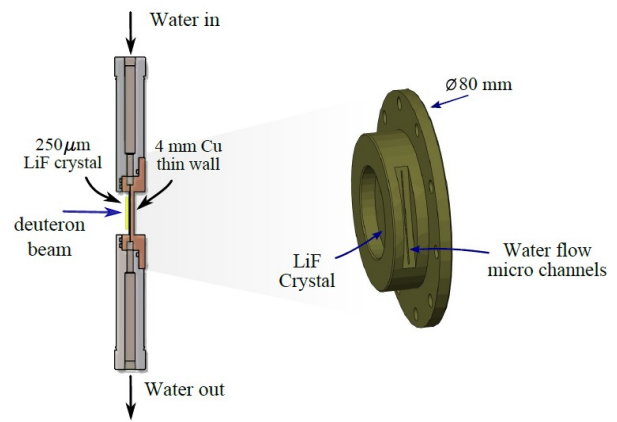

**Figure 15.** A CAD drawing of LiFTiT. Left: Transverse cross section. Right: 3D depiction of the LiF crystal and the copper back. The water micro channels are illustrated as a series of short vertical lines on the element side.

The activation foils data was unfolded by using the cross section evaluations for each foils as extracted from [86] and consequently using a 'guess' spectrum from MCUNED and a 'modified guess' spectrum from the LS data. The unfolded neutron spectrum for both guess spectra are given in Figure 16. The neutron energy reaches up to 20 MeV, as expected from the high Q-value of the neutron generating reactions, and the peak around 13 MeV, which is predicted by MCUNED, is visible as well. The difference between the two spectra, along with the uncertainties in the respective cross sections and gamma activation values, leads us to estimate a ~20% uncertainty for the neutron yield values.

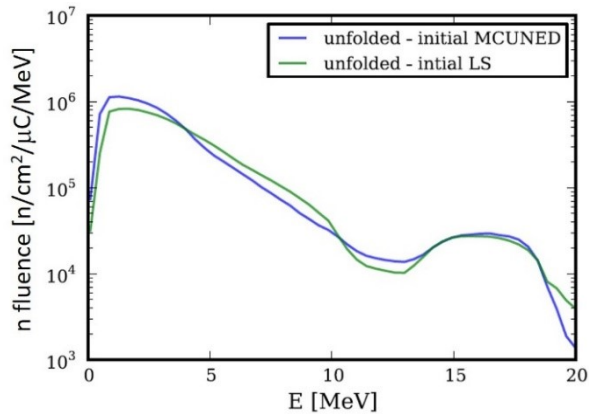

**Figure 16.** Unfolded spectrum of LiFTiT by using different initial guess spectra. Spectra are normalized to reflect the neutron spectrum at 6 cm distance from the target.

Following this successful proof of principle, we plan to repeat these experiments with more intense deuteron beams, in order to test a $^8$Li production apparatus that is based on LiFTiT. The apparatus includes a secondary production target made of 65% porous $B_4C$ disks, mounted inside a high temperature furnace. $^8$Li, produced via $^{11}B(n,\alpha)^8$Li, diffuses out of the $B_4C$ target and is ionized in a thin rhenium surface ionizer. The ionizer and $B_4C$ cage are biased to several kV so the extracted $^8$Li are accelerated towards a detection station where the efficiency of $^8$Li creation will be measured by alpha and beta measurements. For further details, see [82, 37].

### 3.5 Deuteron Induced Cross Sections

The availability of low energy deuteron beams (2.7 – 5.6 MeV) at SARAF I prompted us to perform a series of cross section measurements on several materials. Deuteron cross section data is scarce compared to data on reactions of protons or alpha particles. This data is important for assessment of the activation of components of Radio Frequency Quadrupole (RFQ) injectors and Medium Energy Beam Transport beam dumps in modern deuteron LINACs. We note that the energy range below 5 MeV is especially important as the highest beam loss for these accelerators occurs in this range.

Deuteron cross section measurements are usually performed in cyclotrons using the stack-foil technique. In this method as the beam penetrates through a stack of thin foils its energy decreases. However, as the beam loses its energy in the stack, its energy spread increases significantly, reducing the accuracy of the obtained results especially in the energy range below 5 MeV.

For accurate measurements in this low energy region, a special target station enabling fast target sample replacement was built. Thin targets, transparent to the beam, were used in these experiments. Beam passed through a target with little energy degradation and was collected in the beam dump. Each target was irradiated at a specific energy ($\sigma \sim 25$ keV) and was then taken to a low-background high purity germanium detector station for activation measurements.

The first experiment was for measuring natural copper activation, as copper is the dominant material in RFQ injectors. We measured the $^{63}Cu(d,p)^{64}Cu$ and $^{nat}Cu(d,x)^{65}Zn$ cross-sections in the 2.77–5.62 MeV energy range, and obtained in this energy range results which are consistent with previous measurements, but with much better accuracy, especially in energy [87].

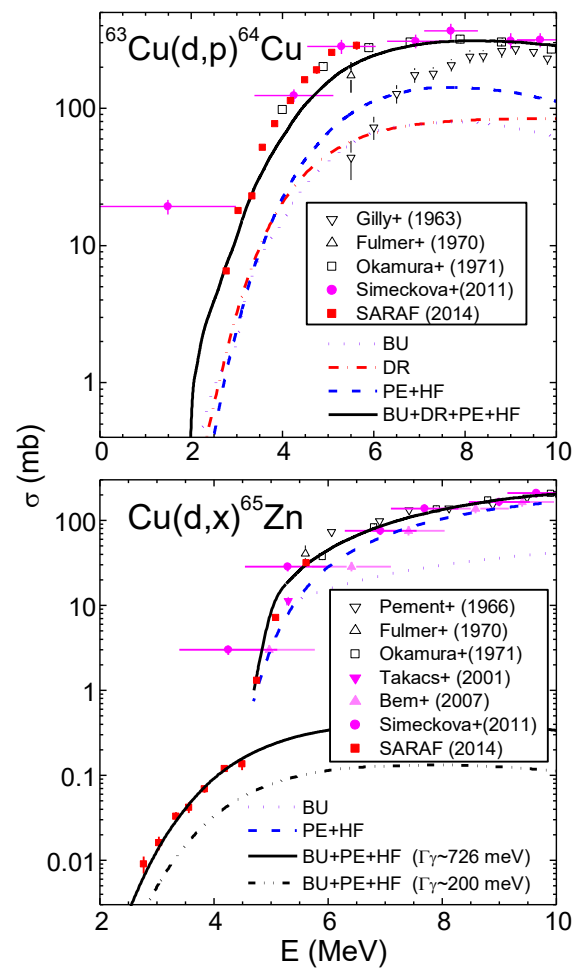

**Figure 17.** Top - Cross section of $^{63}Cu(d,p)^{64}Cu$. Dotted line: deuteron inelastic-breakup (BU), dot-dashed line: direct reaction (DR) stripping component, dashed line: pre-equilibrium (PE) + compound nucleus Hauser-Feshbach (HF) corrected for deuteron-flux leakage through the breakup and DR processes, solid line: sum of all of the above. Symbols: experimental data [87, 88, 89, 90, 91]. Bottom – Cross section of $^{nat}Cu(d,2n)^{65}Zn$ (4.5 MeV and above) and $^{nat}Cu(d,\gamma)^{65}Zn$ (below 4.5 MeV). Dotted and dashed lines are defined as above. Solid line is without direct reaction and corresponds to the average s-wave radiation width $\Gamma_\gamma$ value of ~726 meV [92] for $^{65}Zn$, while the dash-dotted line is related to $\Gamma_\gamma \sim 200$ meV [89]. Symbols: experimental data [87, 88, 90, 91, 94, 95, 96].

The theoretical description of deuteron induced reactions is challenging due to the weak binding energy of the deuteron, $B_d = 2.224$ MeV, which causes a complex interaction process that involves a variety of reactions initiated by the neutrons and protons due to deuteron breakup. Nevertheless, a comparison between $^{63,65}Cu(d,xp)$, $(d,xn)$ experimental data at $E_d = 4$-20 MeV and extensive calculations incorporating several mechanisms resulted in overall agreement [88].

These calculations were recently extended to the lower energy region of the SARAF-I results. The experimental data and theoretical calculations are shown in Figure 17. It is clearly seen that the sum of all mechanism contributions, i.e. deuteron breakup (BU), direct reaction (DR), pre-equilibrium (PE) and compound-nucleus Hauser-Feshbach (HF) calculations describes well the $^{65}Cu(d,2n)^{65}Zn$ data as well as the $^{63}Cu(d,p)^{64}Cu$ reaction particularly due to the DR consideration.

As described in [87], the data points below 4.5 MeV in the bottom plot of Figure 17 are attributed to $^{63}Cu(d,\gamma)^{65}Zn$, and there one may observe a discrepancy with the calculations, if the radiative strength function (RSF) of the compound nucleus $^{65}Zn$ would be assumed according to the former data systematics within the mass range A=61-69 [89] and the related data of $^{64,66}Zn$ nuclei. However, a RSF normalization to the RIPL3 [92] value of the average s-wave radiation widths $\Gamma_\gamma$ is fully supported by the differences found most recently [97] between the RSF data of the rather similar $^{64,65}Ni$ nuclei, and leads to a good agreement with the SARAF-I results for this reaction.

The second experiment was measurement of the $^{23}Na(d,p)^{24}Na$ cross-section in the energy range 1.7–19.8 MeV [98]. At energies below 5.5 MeV, the measurements were performed at SARAF-I using the technique described above for the copper experiment. Above 5.5 MeV, the measurements were performed at the U120M cyclotron in NPI ASCR, Rez. The targets in both sites were thin NaF foils.

The main motivation of this measurement was to establish $^{23}Na(d,p)^{24}Na$ as a standard monitoring cross section in future measurements of deuteron reactions. A secondary motivation is to resolve literature ambiguities in deuteron induced activation of Na, which is the most abundant impurity in lithium, as part of LiLiT development program (Section 2.2.1).

We note that energies down to 1.68 MeV, below the minimal RFQ exit energy of 3 MeV, were achieved by setting the second superconducting cavity downstream the RFQ at a deceleration phase, and using a 'mini-stack' of three NaF foils, with 12 μm aluminum degraders between them. This minimal degradation did not have a significant effect on the energy spread.

Subsequent deuteron cross section experiments were further performed at SARAF-I. The analysis of recent measurements on vanadium and cobalt targets in the 2.7-5.5 MeV energy range was recently completed and is being submitted for publication [99].

## 3.6 Material Science and Radiation Damage Measurements

Radiation damage from proton irradiation exhibits specific features due to hydrogen retention in the metal, such as hydride formation, embrittlement, and nucleation and growth of hydrogen bubbles [100, 101]. Over the last two decades, proton radiation damage in tungsten and its alloys has been of increased interest, due to its selection as a structural material for nuclear fusion reactors, exposing it to low energy proton plasma and a high temperature environment [102, 103, 104, 105, 106].

A blistering threshold dose for tungsten implanted with H (or D) has been reported to lie in the range of $10^{18}$–$10^{20}$ protons/cm$^2$ [103, 107, 108]. This quantity is a critical design parameter for tungsten based accelerator facilities components, both for construction and maintenance.

For several years, we are studying at SARAF-I the blistering thresholds of tungsten irradiated by several MeV protons as a function of target temperature and proton fluxes. Initial results of this study are presented in Figure 18. It has been shown that the critical threshold for tungsten blistering by MeV protons is lower by at least an order of magnitude than reported in the literature for keV proton irradiation. The blisters formed at the MeV proton irradiation conditions are significantly larger than those formed at keV and their morphology depends on the irradiation temperature/flux. For further details on this study, see [55].

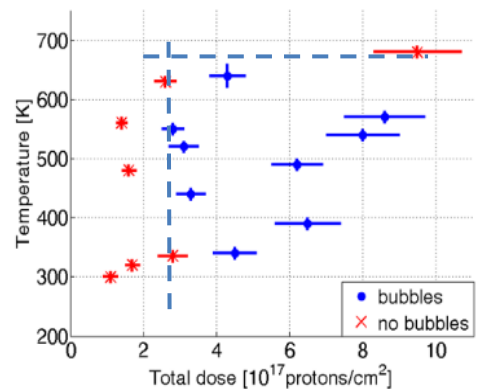

**Figure 18.** Formation of bubbles in tungsten as a function of proton irradiation temperature and total dose. Dashed lines suggest possible boundaries of bubble formation. The total dose uncertainty was derived from beam current and transverse size uncertainties. Reprinted with permission from [55].

## 3.7 Radiation Dosimeters Response to Protons

LiF:Mg,Ti (TLD-100) tags are usually used for dosimetry of gamma radiation. The response of this material to protons was measured at SARAF I, as part of a research program to understand the effects of heavy charged particles (HCP) on TLD-100 within track structure theory (TST).

Irradiations at SARAF were carried out on TLD-100 samples at a fluence of $1.9 \times 10^{11}$ protons /cm$^2$ at beam energy of 1.37 MeV, as part of a series of irradiations at several facilities using different beam particles and energies. This study yielded important insights regarding the treatment of HCP response in TST, and is described in detail in [109].

Another study explored using CR-39 solid-state nuclear track dosimeters for beam halo monitoring at SARAF. Beam halo is a major source of residual activation of the accelerator structure and concrete shielding, and given SARAF's beam current, might limit our ability to fulfill the top-level requirement of hands-on maintenance (Table 1).

CR-39 detectors are suitable for this purpose due to their very high sensitivity to halo particles that hit them, without being affected by the much higher current that passes near them in the core beam. Preliminary tests at SARAF showed feasibility of beam halo measurements for a proton beam of ~2 MeV down to a resolution of a single proton [110].

# 4 SARAF-II Research Plans

## 4.1 The Uniqueness of SARAF-II

Irradiation of deuterons on light targets in the range of tens of MeV, as planned at SARAF-II, will generate fast neutrons with a forward flux and energy peaked around the beam energy divided by 2.5. It will thus be possible to obtain with SARAF-II local fluxes of high energy neutrons that are significantly higher than those available at spallation sources. For example, irradiation of lithium by 40 MeV deuterons generates a local neutron flux in the energy range $5 - 25$ MeV, which is higher by an order of magnitude than that obtained by irradiation of 1,400 MeV protons on tungsten (Figure 19). Notice that the comparison is per $\mu$C, so SARAF-II's advantage is even more significant given its beam current is higher by three orders of magnitude with respect to ISOLDE's 2 $\mu$A proton beam [111].

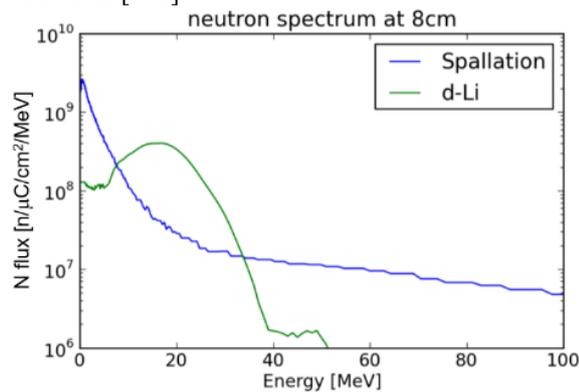

**Figure 19.** Neutron spectra, simulated for 40 MeV deuterons on Li, and measured for spallation 1,400 MeV protons on W (ISOLDE), 8 cm downstream of the targets at 0°.

In Table 7 we compare the expected neutron output of SARAF with the similar upcoming medium energy deuteron projects, SPIRAL-II and IFMIF. In all three projects, the plan is to accelerate deuterons to 40 MeV, but whereas the current at SPIRAL-II and SARAF II will be up to 5 mA, IFMIF is designed as a grouping of two

125 mA accelerators. SARAF-II and IFMIF plan a liquid lithium target, and SPIRAL-II a solid carbon target. In SARAF-II estimations we assume a lithium jet flow of 20 m/s and a flattened transverse beam distribution.

These current-target combinations lead to SARAF-II having the smallest possible beam-spot on target (4th row in Table 7), and therefore the highest beam density on target (5th row). This in turn gives SARAF a maximum neutron flux that is ~2.5 times higher than SPIRAL-II, and only a factor of ~4 lower than that of IFMIF (8th row), even though the latter facility's current is 50 times higher.

The flux (8th row) and mean energy (9th row) are simulated at θ=0° on the target back-side, namely the downstream edge of the neutron producing target, which is ~20 mm thick for lithium and ~5 mm thick for carbon. The mean energies are different due to the different target materials.

The very high forward neutron flux marks the uniqueness and competitiveness of SARAF, as it enables efficient neutron irradiation of small and rare targets, or investigation of low cross section reactions. Specific examples of the utilization of this exceptional characteristic are given in the following Sections.

**Table 7.** Anticipated neutron output of SARAF, compared to SPIRAL-II and IFMIF [112]. Details about the various parameters are given in the text.

| Parameter | IFMIF | SPIRAL-II | SARAF-II |
|---|---|---|---|
| Reaction | d + Li | d + C | d + Li |
| Projectile range in target [mm] | 19.1 | 4.3 | 19.1 |
| Maximal beam current [mA] | 2×125 | 5 | 5 |
| Beam spot on target [cm$^2$] | ~100 | ~10 | ~2 |
| Beam density on target [mA/cm$^2$] | 2.5 | 0.5 | 2.5 |
| Neutron production ratio over $4\pi$ (n/d) | ~0.07 | ~0.03 | ~0.07 |
| Neutron source intensity (n/sec) | ~$10^{17}$ | ~$10^{15}$ | ~$10^{15}$ |
| Maximum neutron flux on target back-side (0-60 MeV) [n/sec/cm$^2$] | ~$10^{15}$ | ~$10^{14}$ | ~$2.5 \times 10^{14}$ |
| <En> on target back-side [MeV] | ~10 | ~12 | ~10 |

## 4.2 Nuclear Astrophysics

### 4.2.1 Extended Measurements with Epi-Thermal and Fast Neutrons

As described in Section 3.1, studies of the astrophysical s-process via epithermal neutron capture reactions constitute the main research program of SARAF Phase I. These studies may continue in the completed SARAF, and in addition, the increased flux of neutrons at higher energies will enable further studies of astrophysical phenomena.

According to estimated temperatures at the various s-process sites, experimental MACS values are required for neutrons with energies of $k_B T = 8$ to 90 keV, which means that neutron capture cross sections should be known up to several hundred keV. Temperatures at r-process nucleosynthesis sites are typically ≤1 GK, corresponding to $k_BT$ values of at most 90 keV as well. However, the p-process operates at higher temperatures, with $k_BT$ values up to 270 keV, requiring neutron cross sections up to at least 1 MeV [113].

Specific neutron induced reactions that gained much interest in recent years are those that destroy $^{26}$Al in massive stars ($^{26}$Al(n,p)$^{26}$Mg and $^{26}$Al(n,$\alpha$)$^{23}$Na) at temperatures up to ~10 GK (equivalent to $k_BT \sim 900$ keV) [114]. $^{26}$Al has astrophysical significance because it can be galactically mapped via its 1809 keV $\gamma$ line, thus providing a 'snapshot' time-integrated view (over a scale of $t_{1/2}(^{26}$Al$) \sim 720$ ky) of ongoing nucleosynthesis in the Galaxy [115]. In particular, its galactic abundance infers the frequency of core collapse supernovae in the Galaxy.

Nevertheless, the available data are sparse, largely because $^{26}$Al is radioactive. As detailed in Ref. 114, data for these reactions is available up to $E_n \sim 100$ keV, and at higher energies it is inferred from the p and $\alpha$ induced inverse reactions.

The high neutron flux at SARAF will enable direct neutron measurements at all relevant energies even on rare, radioactive and extremely small targets such as $^{26}$Al.

### 4.2.2 Study of Neutron-Rich Fission Fragments and Relation to the r-Process

Detailed studies of neutron-rich nuclides provide crucial input to the study of the astrophysical r-process [116]. Such nuclides may be generated at SARAF via neutron induced fission on thin $^{238}$U targets [117]. This reaction is optimal for generation of neutron-rich nuclides at $E_n \sim 10$ MeV [118]. A prototype for generation and extraction of high energy neutron induced fission products (FPs) has been demonstrated at the IGISOL facility at U. Jyvaskyla [118], based on a solid Be neutron converter, which is suitable to their beam current of 10's of $\mu$A.

For the expected 5 mA beam current at SARAF, one should utilize a liquid lithium target, similar to the one that is designed for the thermal neutron source (Figure 5). Further, in quest of maximum stopping and extracting efficiency and extraction times of a few tens of milliseconds, we envision a system inspired by the Ion Catcher at the Fragment Separator (FRS) at GSI [119]. Such an Ion Catcher comprises a Cryogenic Stopping Cell (CSC), which will thermalize the FPs and extract them fast, a Radio Frequency Quadupole (RFQ) that will transport and coarsely mass separate them, and a Multiple-Reflection Time-of-Flight Mass-Spectrometer (MR-TOF-MS) with isobaric (and even isomeric) separation and identification capabilities. A possible schematic layout of the SARAF Ion Catcher is depicted in Figure 20.

The neutron flux from the reaction $^7$Li(d,xn) at $E_d = 40$ MeV is forward peaked, with a relatively wide energy distribution around ~17 MeV (Figure 19 and [38]). Due to technical constraints, the minimal distance between the lithium jet back-side and the internal $^{238}$U thin target is estimated at ~10 cm. A reasonable portion of the neutrons are generated within 0.4 sr, so the transverse area of the target should be 40 cm$^2$. Notice that in principle, one can double the FP rate by adding a second thin target near the downstream wall of the CSC (Figure 20). Additional targets in the CSC bulk are also possible, but they might block the FPs from the former targets, so this needs to be studied in detail. In any case, in the forthcoming rate estimates we use only the first thin target.

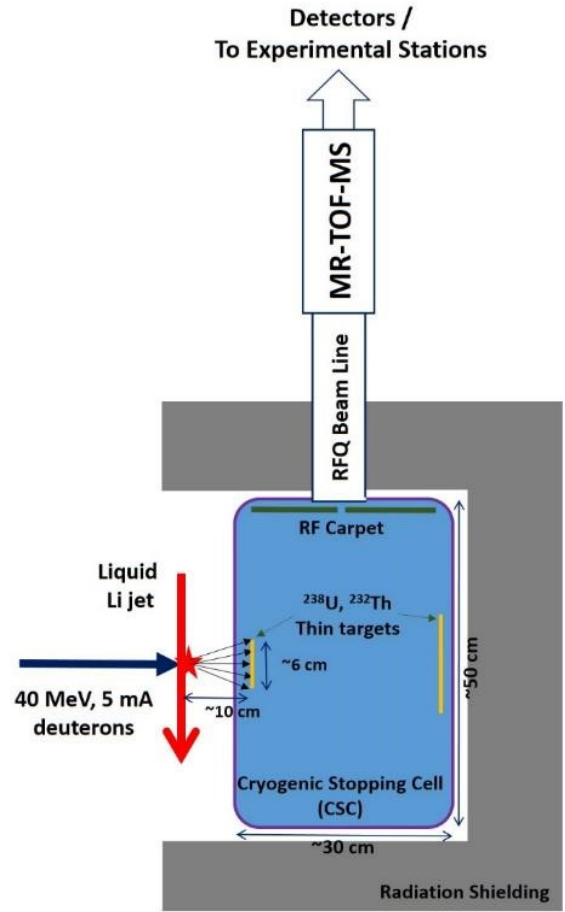

**Figure 20.** Schematic layout of a neutron-induced fission fragment generation, extraction, identification and research facility at SARAF. Other than the CSC, the lithium-target distance and the targets transverse dimensions, the drawing dimensions are not to scale. The overall footprint may decrease by using curved RFQ ion guides and a post-analyzer deflector downstream of the MR-TOF-MS [126]

In order to avoid direct irradiation of the RFQ and MR-TOF-MS instrumentation by SARAF's high neutron flux, we plan to use right-angle extraction from the CSC, a feature that is anticipated in design of the future CSC for the Super-FRS at FAIR [120]. The helium buffer gas density in this future CSC is expected to be ~0.2 mg/cm$^3$ [120], which corresponds to an average FP range of ~10 cm (derived from SRIM2013 [121], based on ~100 MeV per FP).

The CSC will thus be a rectangular box, with a depth of ~30 cm and transverse dimensions of ~50×50 cm$^2$, which will enable almost full stopping of FPs. Extraction will be performed by electric fields that will guide the FPs towards the exit (top flange in Figure 20). In order to avoid FP losses in the top flange, a repelling RF field is required. Given the buffer gas high density, one requires very small electrode structures that can be generated only with a printed circuit board, nicknamed 'RF Carpet' [122, 123]. The extraction efficiency towards the RFQ beam line is expected to be ~60% [124].

The RFQ beam line and MR-TOF-MS will be based on instruments that are experimentally demonstrated. The RFQ beam line may be ~1 m long, and its expected transmission will be ~90% [125]. The MR-TOF-MS may be of similar length, and its expected efficiency will be ~50% [126]. We note that incorporation of curved RFQ ion guides and a post-analyzer reflector downstream of the MR-TOF-MS [126] may decrease the overall footprint of the system.

The FPs will be measured with a sub-nanosecond TOF detector, which detection efficiency is ~80% [127], enabling mass spectroscopy with resolving power of a few $10^5$, accuracy of ~$10^{-7}$, and isobaric and isomeric separations with intensity ratios of up to a few hundred [126, 128], characteristics that are sufficient for most nuclear- and astro-physics needs. Conversely, they may be transported to further experimental stations (with an estimated transmission of ~80% as well), which can include an array of neutron, beta and gamma detectors such as VANDLE [129], and/or a laser spectroscopy system such as the one at the IGISOL facility at U. Jyvaskyla [130].

We now proceed to estimate the expected FP rate at the SARAF n-rich facility. The $^{238}U$ neutron-induced fission cross section at $E_n \sim 15$ MeV is ~1.2 barn [86]. Integrating over the neutrons angular distribution yield [38] up to 0.4 sr, assuming a 5 mA deuteron beam, and a 10 micron thick $^{238}U$ target that covers this solid angle (40 cm$^2$, 760 mg), the estimated fission rate inside the $^{238}U$ target is ~1.6×$10^{11}$ fissions/sec.

Of the above fission products, about 0.5 are expected to escape the $^{238}U$ target [131], of which only 0.5 will be towards the bulk of the CSC, as the target will be installed in its upstream edge. The FPs are distributed initially according to their independent fission product yields (IFY). We used the most updated JENDL data base (March 2012), which was prepared based on ENDF/B-VI data with several modifications [132].

For lower-N fission products, the neutron induced IFY from $^{232}Th$ is higher than that of $^{238}U$, and for a significant amount of cases it compensates the factor of ~3 lower $^{232}Th(n, fiss)$ cross section [86]. Therefore, we assume that we will be able to replace $^{238}U$ and $^{232}Th$ thin targets in the CSC, so for each isotope we use the maximal production rate between these targets (with the same areal density). Further, based on the expected neutron energy distribution (Figure 19 and [38]), for each isotope we used an IFY that is a weighted average of the values at 14 MeV (60% weight) and 0.5 MeV (40% weight), which are the only relevant energies where data is available [132].

Taking into account all the efficiency and transmission factors that are described in the previous paragraphs, we obtain the rates for the isobarically separated FPs that will be detected in the TOF detector downstream of the MR-TOF-MS, or transferred to the further experimental stations. These rates are shown in the top panel of Figure 21. The lowest shown values are $10^{-4}$ isotopes/sec, determined by the minimal IFY values that are given in the database [132].

To put the values of the top panel of Figure 21 in perspective, we compare it to the expected rates of the same isotopes at the future Facility for Rare Ion Beams (FRIB), currently under construction at Michigan State University [135]. FRIB rates for stopped beams were extracted from [136], choosing the expected values at the completed facility, named 'ultimate FRIB yields' [136]. FRIB yields are based on the LISE++ [137] 3EER [138] model for in-flight fission products cross sections. In the bottom panel of Figure 21 we show the ratios between the expected SARAF rates (as defined for the top panel of Figure 21) and FRIB rates (stopped) [136], for the isotopes whose SARAF rate is higher.

From Figure 21 it is clear that SARAF will provide higher rates than FRIB at the very neutron-rich fission products region, especially in the regions 'north-east' of the (Z=28, N=50) and (Z=50, N=82) doubly-magic points, where the SARAF/FRIB ratio reaches 3-4 orders of magnitude. As can be seen by the black solid lines in Figure 21, a significant amount of these isotopes are along the estimated r-process 'path' or 'flow' [116], which is mostly beyond the reach of current facilities. Further, the region shown in Figure 21 covers a significant portion of the isotopes that are most impactful on the r-process, which are depicted in Figs. 13-16 of [139].

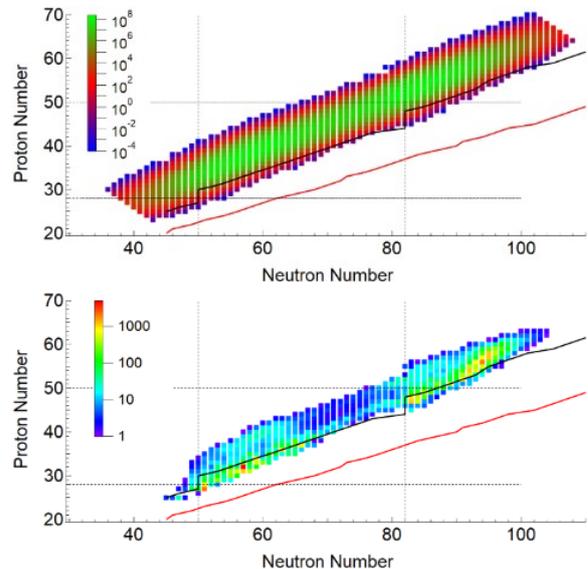

**Figure 21.** Top: Rate of fission products (isotopes/sec) in the TOF detector or the further experimental stations, following extraction and separation in the SARAF Ion Catcher. The maximal rate between $^{238}U$ and $^{232}Th$ is shown. The black line indicates a possible r-process path, defined by $S_n \sim 3$ MeV [133] from the ETFSI-Q mass model [134]. The red line indicates a possible neutron drip line, defined by $S_n \sim 0$ from the same mass model. Bottom: Ratios between neutron-rich nuclide rates at SARAF (as defined for the top panel) and FRIB (stopped), for the nuclides whose SARAF rate is higher. Black and red lines are the same as in the top panel.

These rates may make SARAF the facility of choice for precision measurements of nuclear properties such as masses, half-lives, Q-values, γ-excitation levels, β-delayed neutron emission branching ratios, neutron spectra, spin, dipole and quadrupole moments and mean-square charge radii of the abovementioned isotopes. Moreover, β-delayed γ and/or neutron spectra could impose substantial constraints on the neutron capture cross sections of the decaying isotopes [140], which cannot be measured directly due to their short half-life.

## 4.3 Searches for Beyond-Standard-Model Physics

Precision measurements of the exotic isotopes β-decay parameters (angular correlations, polarization asymmetry, half-life, transition energy and more) may provide signatures for beyond-Standard-Model (BSM) phenomena, including scalar and tensor transitions, time-reversal violation, right handed currents and neutrinos, and deviation from CKM quark mixing matrix unitarity [141]. The β−ν angular distribution for allowed beta decay from an un-polarized nucleus is given by [142]:

$$dW \propto E_e P_e (E - E_0)^2 \left(1 + a_{\beta\nu} \frac{\overrightarrow{P_e} \cdot \overrightarrow{P_\nu}}{E_e E_\nu}\right) \quad (1)$$

where $E_0$ is the energy released in the decay, and $E_e$, $P_e$, $E_\nu$, $P_\nu$ are the energies and momenta of the electron and neutrino, respectively. The first parameter to be measured at SARAF is the β−ν correlation coefficient, $a_{\beta\nu}$. The current experimental data status of the $a_{\beta\nu}$ correlation coefficient for light and medium isotopes is depicted in Figure 22.

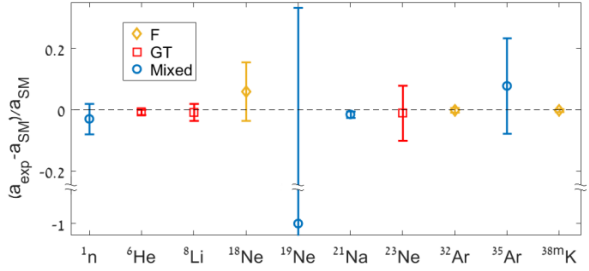

**Figure 22.** Experimentally obtained β−ν correlation coefficients compared to standard-model value for various isotopes. The most precise value was selected for each isotope.

Note that the specific $a_{\beta\nu}$ SM prediction for each isotope depends on its decay type; Fermi, Gamow-Teller (GT), or Mixed. The correlation coefficients affect decay kinematics directly, and since neutrinos cannot be measured, measurement of the beta and or the recoil ion energy distribution provides means of evaluating them.

### 4.3.1 Exotic Isotope Production Scheme

Large quantities of exotic light isotopes will be obtained by a unique two-stage production scheme, which has been introduced in [143] and further studied in [144].

At the first stage, the SARAF 40 MeV deuteron beam impinges on a light nucleus target, specifically our liquid lithium target, to produce a neutron flux with specifications described in Section 4.1. The emerging neutrons irradiate another light material target (the second stage) for exotic isotope production (Figure 23).

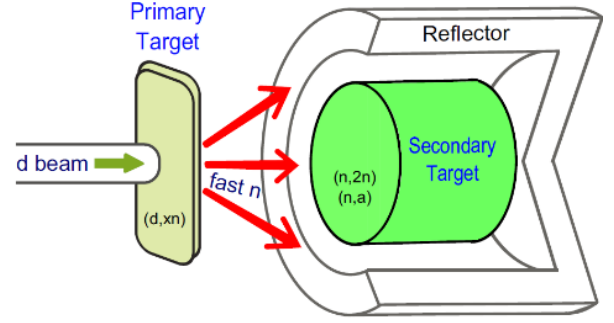

**Figure 23.** Schematic layout of the two-stage production scheme of light exotic isotopes at SARAF. The primary target is a liquid lithium jet, and the secondary target may be porous BeO or B₄C, depending on the researched exotic isotope.

By using a two-stage irradiation we treat the technical problems concerning the primary hot target (heat disposal, radiation damage) and the secondary cold target (diffusion of products) separately, providing easier treatment and maintenance of the entire target system.

The scheme is highly advantageous in terms of production yield since the neutron distribution from the initial break-up spectrum of the deuteron beam on the Li target, matches well the experimental production cross sections of light exotic isotopes. The (unprecedented) expected yields of exotic isotope by this scheme are given in Table 8. The reactions for producing the Ne isotopes at SARAF are given in Table 9. The expected yields for ⁶He, ⁸Li and the relevant Ne isotopes, compared to other existing and future facilities are depicted in Figure 24.

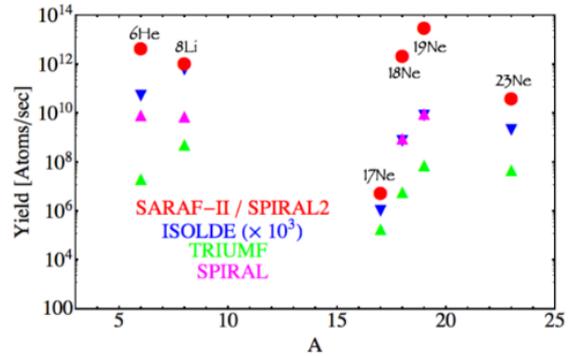

**Figure 24.** Expected yields of light exotic isotopes at SARAF, with a 5 mA proton or deuteron beam, compared to existing and future facilities

**Table 8.** Exotic isotope production yields at SARAF using the two-scheme method described in the text.

| Secondary Target | Reaction | $t_{1/2}$ [msec] | Yield [$10^{12}$ at/mA/s] |
|---|---|---|---|
| BeO | $^9Be(n,\alpha)^6He$ | 807 | 2.5 |
| | $^9Be(n,p)^9Li$ | 178 | 0.033 |
| | $^{16}O(n,p)^{16}N$ | 7,130 | 0.9 |
| B₄C | $^{11}B(n,\alpha)^8Li$ | 838 | 0.87 |
| | $^{11}B(n,p)^{11}Be$ | 13,810 | 0.14 |
| | $^{12}C(n,p)^{12}B$ | 20 | 0.24 |
| | $^{13}C(n,p)^{13}B$ | 17 | 6.6×10⁻⁴ |

**Table 9.** Production reactions of Ne isotopes at SARAF

| Reaction | $t_{1/2}$ [sec] | Threshold [MeV] | Yield [$10^{12}$ at/mA/s] |
|---|---|---|---|
| $^{19}F(p,3n)^{17}Ne$ | 0.109 | 36.8 | $10^{-6}$ |
| $^{19}F(p,2n)^{18}Ne$ | 1.666 | 16.5 | 1 |
| $^{19}F(p,n)^{19}Ne$ | | 4.3 | 10 |
| $^{20}Ne(p,d)^{19}Ne$ | 17.262 | 15.4 | 1 |
| $^{20}Ne(n,2n)^{19}Ne$ | | 17.7 | 1 |
| $^{23}Na(n,p)^{23}Ne$ | 37.140 | 3.8 | 0.02 |

The first SARAF measurements in this context will be of β−ν correlations of $^6$He and Ne isotopes, in search of tensor and scalar components of the β-decay. For $^6$He we will use an Electrostatic Ion Beam Trap (EIBT) [145] and for Ne a Magneto-Optical Trap (MOT) [146].

### 4.3.2 $^6$He measurements using an Electrostatic Ion Trap

In SARAF II we plan to search for beyond-SM signals in the beta decay of $^6$He using an Electrostatic Ion Beam Trap (EIBT) [147]. The advantage of an EIBT in the current context is high detection efficiency due to ion kinematics as well as a large free field region that simplifies the detection configuration. The EIBT is versatile enough and mass independent to be applied in many other cases in which detection efficiency plays a major role. Moreover, the relatively long storage times (few seconds) are ideal for studying isotopes such as $^6$He, having a half-life of 807 ms.

$^6$He decays via beta-minus to $^6$Li and an anti-neutrino, with decay energy of 3.5 MeV. For a pure Gamow-Teller (GT) decay as in this case, the expected angular coefficient is $a_{\beta\nu} = -1/3$. Any deviation from the prediction of -1/3 may indicate physics beyond the standard model.

$^6$He will be generated via WISETRAP (Section 3.3.1). The $^6$He atoms produced in hot target will be purified, transported and injected into an electron beam ion trap (EBIT) [148], where they will be ionized, accumulated, and bunched. The ion bunch will then be accelerated up to ∼4.2 keV, steered, focused, and injected into the EIBT, where beta-decay studies are performed. The beam-line is maintained at a pressure of $10^{-7}$-$10^{-8}$ mbar, and the EIBT chamber is pumped from below by a cryogenic pump to ∼$10^{-10}$ mbar.

The EIBT, which layout is shown in Figure 25, acts similarly to an optical resonator, where ions are redirected between electrostatic mirrors [145]. Each mirror consists of eight electrodes that produce a retarding field that reflects the beam along its path and focuses it in the transverse plane. The innermost electrodes are grounded to create a field-free region, where the ion beam has a well-defined kinetic energy and direction in space. A pickup ring electrode located inside the trap is used to determine the $^6$He bunch position at any given moment.

Following $^6$He decay, the electron and recoil ion are measured in coincidence, providing all kinematic information necessary to deduce $a_{\beta\nu}$. The position and time of flight, and thus the energy of the recoil $^6$Li, are measured by a micro-channel-plate (MCP) detector with a 6.4 mm center hole inside the EIBT. The position and energy of the electron are measured by a specially designed large area position sensitive detector, comprising a plastic scintillator and a several photomultipliers [149], placed outside of the trap.

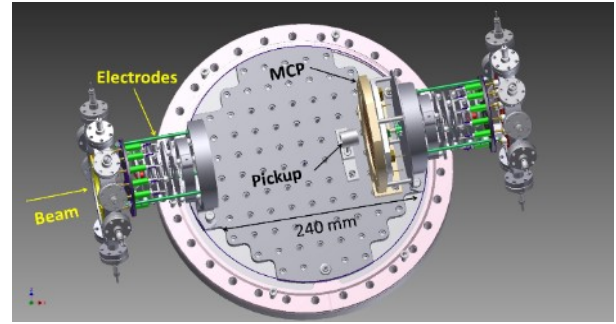

**Figure 25.** Top view of a 3D CAD model of the EIBT. A pickup used for measuring the trapped bunch and a position-sensitive microchannel plate detector are visible in the drawing. Reprinted with permission from [73].

The trap was tested at WIS for various stable ions and exhibited storage times of 0.57 s, 1.13 s and 1.16 s for $^4$He$^+$, CO$^+$ and O$^+$ ions respectively [73]. Trapping of $^6$He ions will be done after analysis of the system uncertainties, installation of the β-detector, and RF bunching in the trap.

### 4.3.3 Ne Isotopes Measurements Using Magneto Optic Traps

The SARAF Magneto Optic Trap (MOT) research program will initially focus on the Ne unstable isotopes due to various physics and technical reasons. As stated in Section 4.3 with regard to Figure 22, scant experimental data for the above Ne isotopes exist, which makes a program focused on them especially attractive. In addition, these Ne isotopes can be produced in very high quantities at SARAF (Table 9), are trappable (in an excited state) in a MOT, their β-decay has not been previously measured in a MOT, and the theoretical calculation of the measured parameters is relatively attainable. Each of the SARAF accessible Ne isotopes has special interesting nuclear properties and further provides access to specific beyond-SM physics scenarios.

$^{17}$Ne is a 2-proton Borromean halo nucleus, which β-decays to the halo nucleus $^{17}$F and its β-decay may be used for searching a non-zero Fierz term and forbidden decays.

$^{18}$Ne decays to the 1.701 MeV state of $^{18}$F in pure GT decay ($a_{\beta\nu}$ = -1/3) and to the 1.042 MeV state in pure Fermi (F) decay ($a_{\beta\nu}$ = 1). Tagging the emitted gamma-ray from the excited state of $^{18}$F will allow simultaneous measurements of both the F and GT decays, constraining the beyond-SM phase space in two orthogonal directions. Further, the decay to the 1.042 MeV state is super-allowed, enabling deduction of $V_{ud}$, which uncertainty for $^{18}$Ne is relatively high [150]. In addition, the transition to the 1.080 MeV state is of interest as it is parity-non-conserving (PNC), and is enhanced by mixing with the 1.042 state [151].

$^{19}$Ne decays almost exclusively to the ground state of $^{19}$F. The ground state decay is a 1/2$^+$ to 1/2$^+$ mirror

transition, which allows for an extraction of the $V_{ud}$ element in the CKM matrix [152]. The $a_{\beta\nu}$ coefficient is close to zero in this case, providing enhanced sensitivity to new physics. The $1/2^+$ nature of $^{19}$Ne allows polarization, making it possible to measure the standard model's polarization dependent correlation coefficients A, B and D. Measurements of the A parameter from polarized $^{19}$Ne decay may offer a high sensitivity to search for right-handed currents [153,154].

$^{23}$Ne decays to the ground state are pure GT, so its $a_{\beta\nu}$ may be measured with high precision and compared to the expected SM value.

Trapping of different neon isotopes is accomplished using a single, 640 nm laser, locked in offset from the $^{20}$Ne cooling transition. The offset is determined by the isotope shift of a specific species, which is on the order of a few GHz [155]. For isotopes with nuclear spin, hyperfine splitting lifts the degeneracy on the cooling transition so atoms may de-excite to states which do not participate in the cooling process. Thus, overlapping laser beams are required to re-pump the atoms back to the cooling cycle. [156]. A view of the engineering model of the MOT system developed for this purpose is given in Figure 26.

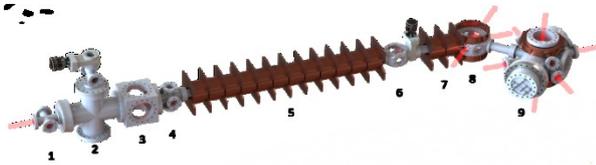

**Figure 26.** CAD model of the MOT setup. 1. Gas input. 2. Metastable source. 3. Collimation stage. 4. Beam diagnostics. 5. Zeeman slower-I. 6. Collimation stage. 7. Zeeman slower-II. 8. Deflection stage. 9. Science chamber

Radioactive neon gas enters the inlet and is excited to a metastable state in the source via an RF discharge using a coaxial resonator [157] with efficiency of about $10^{-5}$ [158]. From that long-lived state there exists a closed transition, where excitation is followed by a decay to the initial state, which is suited for laser cooling [159]. The metastable beam will be collimated directly at the source exit, either by a zig-zag design [160], or white-light molasses [161], before entering a spin flip Zeeman slower [162]. This slows the collimated atoms from thermal velocities of up to 900 m/s to those that can be deflected and trapped (under 50 m/s). The slower is of a novel segmented design, which enables fast, online optimization, rapid tuning of the system for different isotopes, and accounting for stray fields [163, 164].

In order to significantly reduce the background from, and collisions with, ground state radioactive neon, we employ an isotope-selective, forty five degrees deflection-compression stage between the Zeeman slower and science chamber. It has an efficiency estimated at 30%, and utilizes a 1D optical molasses and a magnetic field gradient produced by anti-Helmholtz coils. The compressed beam is deflected and captured in a MOT residing at the center of our large, custom science chamber. It was designed to accommodate three 20 mm MOT beams, three large (40 mm) detectors, additional viewports for trap tomography using CCD cameras, and an electrode setup for collecting recoil charged particles (ions and shake-off electrons) from the trapped cloud with close to 100% efficiency.

The trap lifetime, along with the half-life, determines how many of the trapped isotopes decay within the trap and contribute to our statistics. At the low flow associated with the small amounts of radioactive isotopes produced, collisions with residual background gas at $10^{-10}$ mbar and two-body collisions in the trap play a minor role. Thus intrinsic lifetimes become dominant. For stable neon isotopes, it is the effective metastable lifetime of 15 s [165], which can be extended to up to about 30 s by increasing the excited fraction to the maximum of one half. In these favorable conditions, short-lived neon isotopes (Table 9) will predominantly beta-decay in the trap prior to falling to the ground state, leaving the trap and being pumped out. The longest lived isotope, $^{23}$Ne, will decay within the trap about 40% of the time.

During a decay, singly-charge daughter ions emerge, and due to their low energy (a few hundred eV), virtually all can be collected and detected efficiently with a charged particle detector such as a micro channel plate (MCP). By surrounding the decay volume with a known electric field, the recoil ion's energy spread can be deduced from its time of flight (TOF) distribution. This TOF is measured by triggering on the betas [166], which have high energy and cannot be collected efficiently; or shake-off electrons [167], which might have a low creation probability. A comparison of Monte-Carlo (MC) simulations of ion TOF distributions for different values of $a_{\beta\nu}$ results in its determination and bounds.

The system was commissioned and tested with stable Ne isotopes at the Hebrew University at Jerusalem. Trapping of up to a few $10^6$ atoms at steady-state was recorded using a calibrated CCD camera, and coincidence measurements of recoil ions from collisional ionization

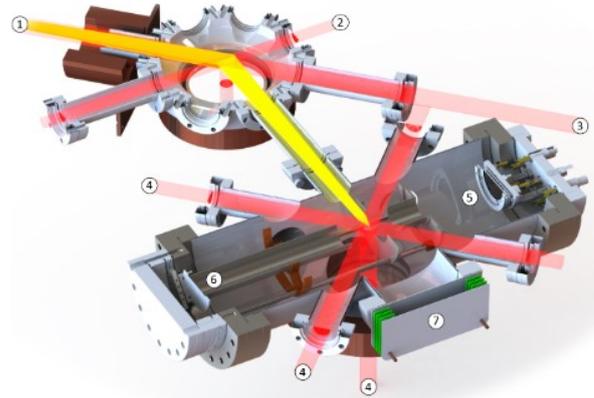

**Figure 27.** Section view of a CAD model of the deflection and science chambers. 1. Slow metastable neon beam (yellow), emerging from the Zeeman-Slower and entering the deflection chamber. 2. Retroreflected deflection laser beam (red), which deflects the neon beam by 45 degrees. 3. Zeeman-slower laser beam (red), for slowing the atomic beam from thermal velocities to about 50 m/s. 4. Three retroreflected MOT beams (red) for trapping. 5. Recoil ion position sensitive detector (PSD). 6. Shake-off electron multi-channel plate (MCP) detector (MCP). 7. Beta multi-wire proportional chamber detector (MWPC) detector.

mechanisms were performed using the collection and detection scheme depicted in Figure 27.

#### 4.4 Nuclear Structure of Exotic Nuclides

In Sections 4.2.2 and 4.3.1 we described the production schemes of exotic fission fragments and light nuclides, with focus on research of the astrophysical r-process and beyond-standard-model physics, respectively. Precision measurements of nuclear properties in exotic isotopes are essential for a better understanding of nuclei and of the underlying interactions between nucleons.

##### 4.4.1 Light Exotic Nuclides

Experimental determination of nuclear properties of light exotic nuclides is especially interesting, because ab-initio calculations provide quantitative predictions of their properties based on empirical nucleon-nucleon and three-nucleon interactions [168].

The measured nuclear properties are mass, spin, parity, nuclear moments, charge radii, lifetime and decay parameters. Most of these properties can be determined by generating ultra-pure samples of exotic nuclei and assembling them in traps, as described in Section 4.3.2 and 4.3.3. It is most advantageous to study a chain of isotopes of the same light element (e.g. He, Li, Be), and compare the results to existing and emerging nuclear models. The dependence on neutron number is compared to ab-initio models and the increase of charge radii for weakly-bound halo nuclei can be observed.

In SARAF-II, it will be possible to study with very high statistics exotic nuclei and unbound states towards the proton drip line up to Z~11 via (p,xn) and (d,xn), and in the neutron-rich region up to Z~4 via (n,γ), (d,p), (n,p), (n,d) and (n,α).

##### 4.4.2 Neutron-Rich Fission Products

Studying the nuclear properties listed in the last paragraph of Section 4.2.2 is important for extending existing nuclear models to the vast region in the nuclide chart between the edge of experimental data and the neutron drip line. A major part of this terra-incognita is covered by neutron-rich fission fragments, which will be accessible at SARAF, as depicted in Figure 21.

In particular, mass measurements in this region can be used for numerous investigations such as studying the limits of nuclear existence, changes in nuclear deformation and onset of nuclear collectivity [169]. Further, masses of neutron rich nuclides impose constraints on models for the nuclear interaction and thus affect the description of the equation of state of nuclear matter, which can be extended to describe neutron-star matter. With knowledge of the masses of nuclides near shell closures, one can also derive the neutron-star crustal composition [170].

Measurements of β-delayed neutron emission probabilities, which is a decay form unique to very neutron rich nuclides, impose constraints on many facets of nuclear physics, from nuclear structure to the reaction mechanism. Future measurements of this property will provide great insight into inner workings of the atomic nucleus and with sufficient data lead to the ability to properly benchmark theoretical models [171].

#### 4.5 High Energy Neutron Induced Measurements

Only limited data exist for high-energy neutron induced reactions, usually at discrete points, and frequently inconsistent with theoretical models, for example in neutron-deuteron breakup reactions [172] and neutron scattering from $^{28}$Si and $^{32}$S [173].

Further, high-energy neutron-induced calibration data is needed for rare event searches. For example, the most significant source of potentially irreducible background in direct dark matter searches are neutrons that scatter elastically from nuclei in the detectors' sensitive volumes, which are mostly made of Ne, Ar and Xe. However, experimental data is available only for a few discrete neutron energies [174].

Another example is the reaction $^{136}$Xe(n,2n)$^{135}$Xe, which is important to guide evaluations and theoretical calculations aimed at estimating the importance of neutron-induced background effects in searches for neutrino-less double-β decay of $^{136}$Xe [175]. Additional high-energy neutron induced reactions important for neutrino-less double-β decay searches are capture on Ge isotopes [176] and inelastic scattering off Pb isotopes [177]. An example of the data status of high energy neutron induced reactions for rare event searches is given in Figure 28. One may notice that above 8 MeV there are measurements only at ~14 MeV, and more measurements in this high energy range are needed.

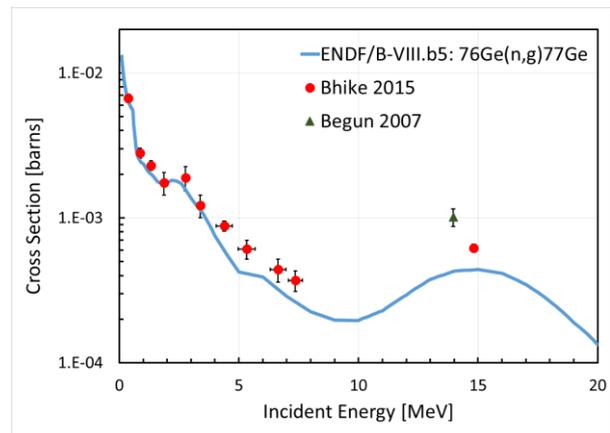

**Figure 28.** Data status of $^{76}$Ge(n,γ). Blue line: ENDF/B evaluation [86]. Symbols: Experimental data from Bhike et al. [176] and Begun et al. [178].

Neutron induced cross sections at high energies are important for the design of Generation-IV fission reactors and nuclear waste transmutation facilities, whose energy spectra are expected to reach ~10 MeV and higher. In particular, there is interest in the neutron induced fission and capture cross sections of long-living trans-uranium (TRU) elements, such as $^{241,242m,243}$Am, $^{243-245}$Cm, $^{237}$Np, and more [179]. Specifically, transmutation to shorter lived isotopes is achieved by neutron induced fission, whereas neutron capture might hinder this process, so their ratio as a function of neutron energy is of interest [179].

Installing these TRU thin targets in the planned SARAF Ion Catcher (Section 4.2.2), in addition to the $^{238}$U and $^{232}$Th thin targets, will enable direct measurements of the fission product distributions of all these isotopes via mass measurements, and also of isomer to ground-state yield ratios. This information is of interest to the IAEA regarding future generation fission reactors, specifically for calculations of post reactor shutdown decay heat [180]. Fission product distributions and isomer yield ratios are also of interest for the prediction of antineutrino spectra generated by nuclear reactors, which is important input to reactor based neutrino oscillations experiments [181].

There are also long-lived fission products (LLFPs: $^{79}$Se, $^{93}$Zr, $^{99}$Tc, $^{107}$Pd, $^{129}$I, and $^{135}$Cs), which pose an additional nuclear waste challenge. It has been recently proposed to explore transmutation via fast neutron capture to shorter lived isotopes [182]. Further, the interplay between neutron capture and (n,2n) reactions may affect the transmutation efficiency [183].

In both transmutation scenarios (TRU and LLFP), neutron induced data in the required energy range is scarce and scattered. An example for the LLFP $^{93}$Zr is given in Figure 29, which shows evaluated (n,γ) and (n,2n) cross sections, the limited direct measurement data, and a relative fast reactor neutron spectrum (in arbitrary units), to demonstrate the importance of measuring (n,γ) and (n,2n) for LLFPs at a wide neutron energy range. We note the existence of indirect $^{93}$Zr(n,γ) data up to $E_n = 8$ MeV, using the surrogate ratio method via $^{90,92}$Zr($^{18}$O, $^{16}$O)$^{92,94}$Zr [184].

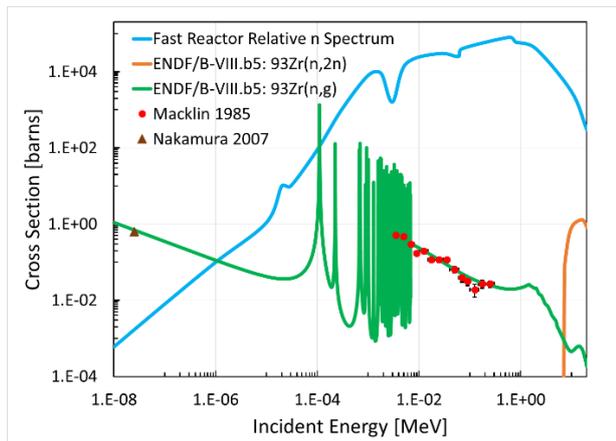

**Figure 29.** Data status of $^{93}$Zr(n,γ) and $^{93}$Zr(n,2n). Green and orange lines: ENDF/B evaluations [86]. Symbols: (n,γ) direct measurements from Macklin et al., [185] and Nakamura et al., [186]. Blue line: Fast reactor relative neutron spectrum (arbitrary units) [182].

An additional need for Generation-IV fission reactors, and also for accelerator driven systems (ADS) and future fusion demonstrators is the study of fast and high energy neutron induced radiation damage. For fusion reactors, one of the main risks is high energy (~14 MeV) neutron induced radiation damage due to d+t fusion. SARAF will have the world's highest local high-energy neutron flux in the near future, expected to be only a factor of ~4 less than the planned IFMIF (Table 7), so it may be considered as an intermediate facility for irradiation of miniaturized material samples (e.g., 0.5 mm beryllium pebbles, which compose the ITER multiplier blankets [187]).

In summary of this Section, SARAF will enable extensive research with high energy neutrons, with a variable energy peak and a maximal value of ~50 MeV, for which data is currently limited, or entirely missing. The high neutron flux of SARAF could cope with the challenges described above, which significant part involve rare and radioactive targets that are available only in minute sizes.

## 4.6 High Energy Deuteron Induced Measurements

The current nuclear data base on deuteron induced cross section is very lacking and there is a demand from research programs like IFMIF, SPIRAL2 and ITER for an improved data base and models for deuteron reactions. The same is true for the SARAF where the role of deuteron-induced reactions in the assessment of the induced neutron yield and the induced radioactivity of accelerator components (elements such as Ga, Al, Cu, Fe, Cr, Nb, etc.) is important.

The lack of the experimental data significantly limits the reliability and validity of the semi-empirical theoretical models calculating deuteron reactions processes, for example TALYS [188]. Frequently, predictions of these models are valid only for the cases for which reliable experimental data are available. Theoretically, deuterons are of special interest due to the importance of the deuteron break up mechanism making them substantially different from other incident particles.

The high deuteron energy, variable of up to 40 MeV, will make SARAF II an excellent facility for differential cross section measurements. The planned setup that we envision will have a gantry with magnets rotating the beam around a thin target, similar to [189]. This will enable (d, xn) double differential cross section measurements in a wide range of angles. The neutrons will travel along a few tens of meters beam line from the target to a matrix of detectors for TOF measurements. With an array of neutron detectors surrounding the target we will perform coincidence measurements to distinguish between (d,n), (d,2n) and (d,3n) reactions, and with a germanium detector (with proper sensitivity, location and shielding to avoid radiation induced damage) we will measure prompt gammas.

This program relies on the SARAF fast beam chopper that is described in Section 2.1.1, with which we already demonstrated ToF measurements via a proton beam with a resolution of σ ~ 1.3 ns (Figure 2). Deuteron induced ToF measurements are ongoing at SARAF I. Further, we will rely on our experience in deuteron induced total cross section measurements that are described in Section 3.5.

## 4.7 Neutron Based Imaging

### 4.7.1 Thermal Neutron Imaging and Diffractometry

Thermal Neutron Radiography (TNR) is a powerful non-destructive imaging technique that is complementary to x-ray imaging. This is because the neutron attenuation in materials is due to nuclear interactions, which behave differently than the attenuation of x-rays by electromagnetic interactions with electrons. Therefore, most light nuclei (e.g. H, B) have high thermal neutron attenuation coefficients, whereas some heavier elements (e.g. Pb) have relatively smaller ones.

TNR is a major application of SARAF II, within its goal to enhance and back up the Soreq IRR-1 5 MW nuclear research reactor [190, 191]. The TNR system at IRR-1 has a neutron flux on the imaging plane of $6 \times 10^5$ n/s/cm$^2$, with a cadmium ratio (that determines the thermal purity of the neutron beam) of ~15, and with a ratio between the neutron collimator - image plane distance (L) and the aperture diameter (D), defined as 'L/D', of 250.

The beam and current top level requirements of SARAF-II (Table 1), in combination with a liquid lithium conversion target, generate a neutron yield of $10^{15}$ n/s. These high energy neutrons, peaked around ~17 MeV (Figure 19 and [38]), are moderated to the thermal energy range (peaked around ~25 meV).

This is achieved by moderating the neutrons in heavy water that is surrounded by a light water reflector, along with a beryllium multiplier that enhances the number of thermal neutrons. Extraction tubes towards envisioned radiography and diffractometry systems are placed at wide angles with respect to the incident deuteron or proton beam, to diminish the contribution of unwanted fast neutrons. The source will be used also for neutron activation analysis (NAA) of samples, which will be transferred into the highest thermal neutron flux region via a pneumatic transfer tube.

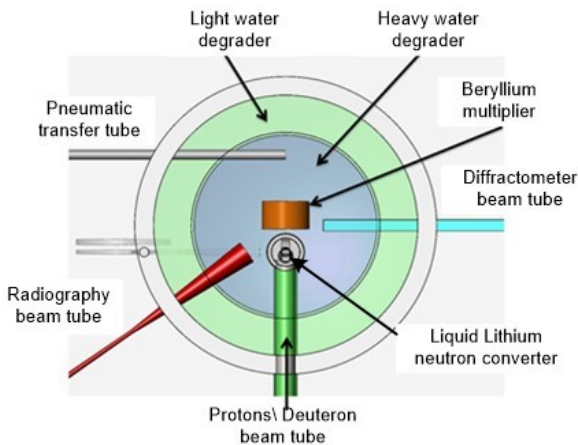

**Figure 30.** Schematic horizontal cross section of the SARAF thermal neutron facility. The liquid lithium neutron converter is downstream of the particle beam tube. Neutrons are thermalized in light and heavy water moderators, multiplied by a beryllium multiplier and then some of them exit towards envisioned radiography and diffractometry systems. A pneumatic tube for transferring samples for NAA irradiation is also depicted.

The exact dimensions and tube locations of the thermal neutron source of SARAF II were obtained by detailed Monte-Carlo simulations (MCNP [191]), and a schematic view of the source is shown in Figure 30.

The simulation results indicate that the neutron characteristics at the imaging plane and at the diffractometer can be similar to those at IRR-1, when irradiating the liquid-lithium target with a 40 MeV, 2 mA CW deuteron beam at SARAF-II.

### 4.7.2 High Energy Neutron Imaging

Reactor based neutron imaging is performed using thermal neutrons (25 meV), due to their intrinsic moderation in the reactor core. However, with an accelerator based neutron source one may use fast neutrons for imaging, which enable good contrast over broad densities and atomic numbers. This is useful for e.g. investigation of dual-phase flow around fuel rods of fission reactors and study of coolant and lubricant flow in combustion engines using fast and slow neutrons.

Preliminary research with 14.2 MeV neutrons from a d-t generator has been performed, studying the effects of different types and materials of collimators on the quality of tomography images. The figure of merit was minimal scattering effects on the resultant image. The conclusions were that one should use fan-beam tomography rather than cone-beam, surrounded by a moderator made of a mixture of Borax (Na$_2$B$_4$O$_7 \cdot$10H$_2$O) and water [193].

## 4.8 Medical Research and Development

### 4.8.1 Accelerator Based Boron Neutron Capture Therapy

Modern neutron based therapy consists mainly of accelerator based Boron Neutron Capture Therapy (BNCT) which is intended for treatment of brain tumors [194]. BNCT is comprised of selectively delivering $^{10}$B to tumor cells and irradiating the tumor region with neutrons. Given the high cross section for $^{10}$B(n,$\alpha$)$^7$Li for thermal neutrons, they are absorbed and the short range (5-9 $\mu$m) $\alpha$ and $^7$Li deliver their dose mainly to the tumor cells.

Suitable neutron sources for BNCT have been limited for many years to nuclear sources, but reactors' thermal neutron spectrum (~25 meV) restricted treatment to superficial tumors. For deep-seated tumors (a few cm), epithermal neutrons (available at accelerators) are favored [195]. Another major advantage of an Accelerator-based BNCT (ABNCT) over reactors is the potential for installation within a hospital.

The use $^7$Li(p,n)$^7$Be with proton energy several tens of keV above the reaction threshold (E$_{thr}$ = 1.8804 MeV) is of special interest for BNCT [196, 197, 198, 199]. It reduces the total neutron yield, compared to higher beam energies, but the maximum and mean neutron energies are much lower as well, requiring less moderation, which leads to less neutron attenuation. Further emitted neutrons have forward kinematics, increasing the usable yield for therapy. Moreover, the lower proton energy (~2 MeV) sharply reduces self-activation of accelerator material, which is of great concern in a high current machine.

In SARAF, the combination of the high CW ion current and the liquid lithium target enables advanced ABNCT research and proof of the technological concept, based on the favored $^7Li(p,n)^7Be$ reaction.

The $^{10}B$ dose created by near threshold $^7Li(p,n)$ neutrons was simulated by Monte-Carlo methods and was measured by irradiating $^{10}B$ samples sandwiched between CR-39 nuclear track detectors, inside a water filled phantom, simulating human tissue. The treatment dose was deduced by counting the $\alpha$ and $^7Li$ tracks produced by in the CR-39 by $^{10}B(n,\alpha)^7Li$ [31].

Based on the technical experience accumulated with LiLiT at SARAF I, a neutron-beam shaping assembly was designed for ABNCT, including a polyethylene (PE) neutron moderator and a lead gamma shield, using Monte Carlo (MCNP and GEANT4) simulations of BNCT-doses produced in the abovementioned phantom. According to these simulations, a ~15 mA near threshold proton current will enable therapeutic doses within a ~1 hour treatment. The experience with LiLiT (Section 2.2.1) indicates that such large currents could be dissipated in a liquid lithium target, deeming this design feasible [200].

The variable energy and increased beam current at SARAF II will enable further research in this field, for possible future application to dedicated compact high current low energy accelerators that could be installed in hospitals.

### 4.8.2 Radiopharmaceuticals Development

Radioisotopes production for radiopharmaceuticals is in growing interest as new diagnostic and treatment procedures are developed. In parallel, the aging of nuclear reactors that supplied most of the world demand put a risk on this supply source (e.g. the $^{99m}Tc$ crisis [201, 202]).

This makes accelerator based radioisotopes production the most viable path to provide this need. Moreover, the versatility of radioisotopes production with accelerators enables development of promising new diagnostic and treatment procedures, which are not available via radioisotopes based reactor production.

SARAF II has the exact specifications that were recommended by the NSAC Isotopes Subcommittee for future radio-isotope production [203]. Its advantages are the variety of attainable radioisotopes due to the variable-energy and particles and the high production yield due to the high current.

In addition to the proper energy range, a viable radioisotopes production program requires appropriate production targets, which can withstand the expected high power and power densities. The research and development efforts described in Section 2.2.2 are directed precisely towards this goal. In the following Subsections we report radioisotopes for nuclear medicine whose development and production at SARAF II has been considered and studied due to the SARAF advantages, mainly for its high deuterons current (Table 10). Few of the examples are discussed in more details herby.

**Table 10.** Radioisotopes for nuclear medicine that have been considered for development and production at SARAF II. The percentage in the $2^{nd}$ column refers to target isotope abundance. Target type is Gas, Solid or Liquid. Energy (E) is in MeV. The half-life ($t_{1/2}$) is expressed in minutes (m), hours (h) or days (d). Clinical usage options are Diagnostics, Therapy and Generator ($^{68}Ge$ generates $^{68}Ga$ and $^{225}Ac$ generates $^{213}Bi$).

| Isot. | Tar. (%) | Tar. type | E | Reac. | $t_{1/2}$ | Clinic. usage |
|---|---|---|---|---|---|---|
| $^{18}F$ | $^{20}Ne$ (90.5) | G | 15 | d,$\alpha$ | 110 m | Diag. |
| $^{64}Cu$ | $^{64}Ni$ (0.9) | S | 21 | d,2n | 12.7 h | Diag. / Ther. |
| $^{67}Cu$ | $^{67}Zn$ (4.0) | S | 14 | n,p | 58.5 h | Ther. |
| $^{68}Ge$ | $^{69}Ga$ (60.1) | L | 30 | p,2n | 271 d | Gen. |
| $^{89}Zr$ | $^{89}Y$ (100) | S | 30 | d,2n | 78.4 h | Diag. |
| $^{99m}Tc$ | $^{100}Mo$ (9.63) | S | 25 | p,2n | 6.0 h | Diag. |
| $^{103}Pd$ | $^{103}Rh$ (100.0) | S | 21 | d,2n | 17.0 d | Ther. |
| $^{111}In$ | $^{112}Cd$ (24.0) | S | 18 | p,2n | 68.0 h | Diag. |
| $^{177}Lu$ | $^{176}Yb$ (12.7) | S | 20 | d,x | 6.7 d | Ther. |
| $^{123}I$ | $^{124}Xe$ (0.094) | G | 30 | p,2n | 13.2 h | Diag. |
| $^{201}Tl$ | $^{203}Tl$ (29.5) | S | 30 | p,3n | 73.0 h | Diag. |
| $^{186}Re$ | $^{186}W$ (29.0) | S | 16 | d,2n | 90.0 h | Diag. / Ther. |
| $^{211}At$ | $^{209}Bi$ (100.0) | S | 28 | $\alpha$,2n | 7.2 h | Ther. |
| $^{225}Ac$ | $^{226}Ra$ (trace) | S | 17 | p,2n | 10.0 d | Ther. / Gen. |

#### 4.8.2.1 Lutetium-177

Lutetium-177 ($^{177}Lu$, $t_{1/2}$ = 6.7 d) is a therapeutic radioisotope with favorable physical characteristics of strong $\beta^-$ emission (497 keV, 78.6%) and $\gamma$-emissions that enable imaging. $^{177}Lu$ is used for radiolabeling of a variety of biomolecules, such as peptides and monoclonal antibodies [203]. The fast growing demand for $^{177}Lu$ in the recent years raises the need for new supply sources.

There are two routes for the $^{177}Lu$ production in reactors: the direct $^{176}Lu(n,\gamma)^{177}Lu$ and the indirect production $^{176}Yb(n,\gamma)^{177}Yb \rightarrow ^{177}Lu$. The direct method is based on thermal neutron irradiation of an enriched $^{176}Lu$ target, a reaction with a high cross section of ~2.9 b [205]. The main disadvantage of this route is that a minimum amount of stable $^{176}Lu$ cannot be avoided, limiting the specific activity of the $^{177}Lu$ in the final product. The requirement for high specific activity, especially for monoclonal antibodies labelling limits $^{177}Lu$ direct production to only a few high flux reactors [206]. In addition, during direct activation of the lutetium target, a long-lived $^{177m}Lu$ radioisotope with a high $\gamma$ energy is produced ($E_{max}$=400 keV, $t_{1/2}$ = 160 days). The presence of this isotope in the final product can cause dosimetric, regulatory, and environmental problems.

In contrast to the direct method, the indirect route produces a non-carrier-added $^{177g}Lu$ (after chemical separation) with high specific activity and the contamination of [177mLu] is eliminated [207, 208]. However, the indirect reaction cross section is much lower, only 2.4 b.

Few attempts to produce [177gLu] in accelerators have been made in the past using hafnium and tantalum targets [209]. However, the need for high energy accelerators (195 MeV), in addition to the low production yield made these methods impractical.

Soreq NRC, in collaboration with the Vrije Universiteit in Brussel, demonstrated the possibility of producing [177gLu] via the reaction of [176Yb(d, x)177Yb,177Lu] [203] with maximum cross section of 200 mb in the energy range of 12-14 MeV. The optimal energy for [177Lu] production is ~20 MeV [203]. We emphasize that the cross section of producing [177Lu] via protons is negligible.

The suitable energy and high deuteron current in SARAF II will enable yields around 1 TBq of [177Lu] in 72 h irradiations with 21 MeV deuterons, at 2 mA beam intensities on enriched [176Yb] targets [203], and is therefore expected to make this production route more cost effective. The irradiation target is usually made by electroplating natural ytterbium on a water cooled copper plate. The preparation of an ytterbium target is fully described in [210].

### 4.8.2.2 Palladium-103

Palladium-103 ([103Pd], $t_{1/2}$=17.0 d) is used for prostate cancer brachytherapy. It is an x-ray (21-23 keV) and Auger electrons emitter. It was originally produced in reactors, but poor production yields and high cost of the enriched [102Pd] resulted in a shift to cyclotron production of a non-carrier-added product by proton or deuteron irradiation of [103Rh] (Figure 31) [211]. Similar to the [177Lu] production by cyclotrons, the production yield of [103Pd] by deuterons is much higher in comparison to the protons route.

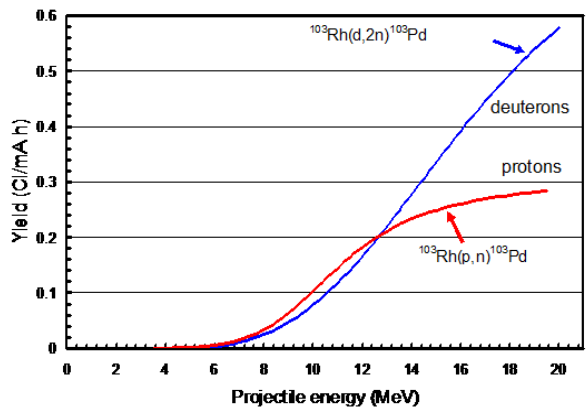

**Figure 31.** Calculated production yields of [103Pd] via proton reaction compered to deuteron reaction on [103Rh].

The irradiation target is made by electroplating of rhodium on a water cooled copper plate similar to the ytterbium target [47]. This irradiation target technique has several inherent drawbacks; first, irradiation is with low beam current due to target cooling limitations, second, the costly and difficult electroplating process used to prepare the target and third, activation of the copper target substrate leads to long-lived radio-contaminations of the target, requiring tedious dissolution and separation procedures. For that reasons, a new design should be developed. In parallel to target design, a palladium/rhodium separation process was developed, based on selective separation of the palladium by an amberjet 4200CL resin [210].

### 4.8.2.3 Germanium-68

Germanium-68 ([68Ge], $t_{1/2}$ = 270.8 d) serves as the parent radionuclide for the preparation of [68Ge/68Ga] generators. The daughter, Gallium-68 ([68Ga], $\beta^+$ 88%, $t_{1/2}$ = 67.6 min), is a short-lived positron emitter utilized in Positron Emission Tomography (PET) imaging. The high potential of [68Ge/68Ga] generators for PET was recognized a few decades ago due to the following reasons: 1) The long half-life of [68Ge] enables design of a generator usable for long periods, and saves the hospital the complexity and expenses of buying and maintaining a cyclotron required for PET radiopharmaceuticals as FDG, 2) The 68 min half-life of the daughter matches the pharmacokinetics of many peptides and other small molecules, and that is the reason for the extensive research on [68Ga]-labelled tracers, 3) Favorable results of recent clinical studies using [68Ga]-labelled peptides [212].

[68Ge] can be produced in accelerators via [69Ga(p,2n)] or [natGa(p,x)68Ge] (natural gallium consists of 60% [69Ga]). Cross section data has been published by Takacs [213], Levkovskij [214] and Cohen [215] and the [68Ge] yield from [69Ga] thick targets was calculated [51]. Accordingly, the proton energy for efficient [68Ge] production is ~30 MeV, with 2440 MBq/(mA·h). Due to the long half-life of [68Ge], high current accelerators with on-target beam intensities in the range of 100 μA are required for efficient production of sufficient batch yields (about 37 GBq)

SARAF II can produce raw material for a single high activity generator within 36 hours, versus an average irradiation time of 100 days in other facilities [216, 217]. The short irradiation time is expected to reduce production costs by an order of magnitude, making this valuable diagnostic and therapeutic agent more affordable. To achieve these high production rates, one should overcome the thermal and mechanical difficulties associated with constructing the required high power irradiation target. A novel target design to enable [68Ge] production by a high current (> 1 mA) proton beam at 30 MeV was developed in SARAF, and is described in Section 2.2.2 and in [218].

### 4.8.2.4 Actinium-225

Alpha particles have several advantages for therapy over beta particles such as (a) high linear energy transfer (LET) (b) short path lengths in tissues (50-80 μm) (c) limited ability of the cells to repair damage to the DNA [219, 220]. Actinium-225 ([225Ac], $t_{1/2}$ = 10.0 d) is used for radioimmunotherapy of cancer (see e.g. [221]). It is an alpha emitting radionuclide (6 MeV) that generates 4 net alpha emitters in a short decay chain to stable [209Bi]. [225Ac] can be used also as a short lived [213Bi] generator (46 min, 6 MeV). [225Ac/213Bi] generators are available commercially with activities of 925 – 3700 MBq.

$^{225}$Ac is the natural product decay of $^{233}$U via $^{229}$Th. However, $^{229}$Th has a very long half-life (t$_{1/2}$ = 7340 y) and therefore, the activity that can be produced in this pathway is very limited worldwide [222, 223]. Hence, the most promising production route of $^{225}$Ac is in accelerators or reactors [222, 223]. $^{225}$Ac can be produced via neutrons, deuterons or gamma-rays, but the preferred method is to irradiate $^{226}$Ra with protons in the reaction $^{226}$Ra[p, 2n]$^{225}$Ac [219] where the optimum energy is 16.8 MeV [223]. The theoretical irradiation yield for production of $^{225}$Ac is 7 kBq/($\mu$A·h)/mg of $^{226}$Ra. With a high current of 2 mA, SARAF II can produce ~30 GBq in 72 h irradiations, where the average activity for cancer treatment is 37 MBq.

*4.8.2.5 Astatine-211*

Astatine-211 ($^{211}$At, $\alpha$ 42%, 5.7 MeV, t$_{1/2}$ = 7.2 h) is used for radioimmunotherapy, mainly for ovarian cancer [224]. $^{211}$At decays also by electron capture (58%) to its short-living alpha emitting daughter ($^{211}$Po, 7.5 MeV, t$_{1/2}$ = 516 msec) that decays to the stable $^{207}$Pb. $^{211}$At is produced in accelerators by irradiation of a natural bismuth target in the reaction of $^{209}$Bi($\alpha$,2n)$^{211}$At at optimal energy of 28 MeV.

In low currents of 50-60 $\mu$A, the production yield is 1520 MBq/$\mu$A·h [225, 226]. In SARAF II, with an additional ion source designed for nuclides heavier than hydrogen, the alpha particles current could reach 2.5 mA, and increase the production rate accordingly.

# 5 Summary and Outlook

SNRC is constructing SARAF-II, a new national scientific infrastructure in Israel that is based on a high-power proton/deuteron accelerator, for basic and applied nuclear physics research, to be completed by the beginning of the next decade.

SARAF-II's cutting-edge specifications, with its unique liquid lithium target technology, can make this facility competitive with the world's most powerful deuteron, proton and fast neutron sources.

SARAF-I is already operational. Ground breaking results in high current superconducting acceleration and high power target irradiation have been achieved. International scientists and students perform forefront research and development at SARAF, which will be expanded via a new target room built for Phase-I [227], to be completed by the beginning of 2018.

We continue to develop the SARAF-II scientific research program, and look forward to letters of intent and proposals, for both SARAF-I and SARAF-II.

We gratefully acknowledge the SARAF technical team at SNRC for development, operation and maintenance of the SARAF accelerator, targets and insfrastructure. We wish to thank Dorothea Schumann and the Isotope and Target Chemistry group at Paul Scherrer Institute (Switzerland) for the supply of radioactive targets and their collaboration. The work presented in this review is supported by grants from the Pazy Foundation (Israel), Israel Science Foundation, German–Israeli Foundation, US-Israel Bi-national Science Foundation, European Research Council, and Minerva Foundation. BO and MT are supported by the Ministry of Science and Technology, under the Eshkol Fellowship. MP gratefully acknowledges support along successive years of Pazy Foundation, German Israeli Foundation (No: G-1051-103.7/2009), Binational Science Foundation (No. 2012098), Israel Science Foundation (No. 2014/1387/15). MG acknowledges support from the U.S.-Israel Bi National Science Foundation, Award Number 2012098, and the U.S. Department of Energy, Office of Science, Office of Nuclear Physics, Award Number DE-FG02-94ER40870.